\documentstyle[preprint,aps,prbbib,epsf]{revtex}
\newcommand{\vct}[1]{\mbox{\boldmath $#1$}}
\newcommand{\lsim}
{\ \raise.35ex\hbox{$<$}\kern-0.75em\lower.5ex\hbox{$\sim$}\ }
\begin{document}
\draft
\title{Tunneling conductance
of normal metal / $d_{x^{2}-y^{2}}$-wave superconductor
junctions in the presence of
broken time reversal symmetry states near interfaces}
\author{Y. Tanuma \cite{Okayama}, Y. Tanaka}
\address{Department of Applied Physics,
Nagoya University, Nagoya 464-8063, Japan}
\author{S. Kashiwaya}
\address{Electrotechnical Laboratory, 
1-1-4 Umezono, Tsukuba, 305-0045, Japan}
\par
\date{\today}

\maketitle
\begin{abstract}
In order to clarify the influence of (the presence of)
the broken time-reversal symmetry state (BTRSS)
induced near the interface,
tunneling conductance spectra in
normal metal / $d_{x^{2}-y^{2}}$-wave superconductor
junctions are calculated
on the basis of the quasiclassical
Green's function method.
The spatial dependence of the
pair potential in the superconductor side
is determined self-consistently.
We discuss two types of the symmetry on the BTRSS;
i) $d_{x^{2}-y^{2}}$+i$s$-wave state and
ii) $d_{x^{2}-y^{2}}$+i$d_{xy}$-wave state.
It is shown that the amplitude of the subdominant component
(i$s$-wave or i$d_{xy}$-wave)
is quite sensitive to
the transmission coefficient of the junction.
As the results,
the splitting of the zero-bias conductance peak due to the BTRSS inducement
is detectable only at junctions with small transmission coefficients
for both cases.
When the transmission coefficients
are relatively large, the explicit peak splitting does not occur
and the difference in the two cases
appears in the height  of the zero-bias peaks.
\end{abstract}
\par
%
\section{Introduction}
Identifying of the
pairing symmetry in high-$T_{c}$ superconductors
is important to clarify the mechanism of the superconductivity.
A great deal of experimental and theoretical
studies have revealed that
the superconducting pair potential
has $d_{x^{2}-y^{2}}$-wave symmetry in the bulk state
\cite{Scalapino,SR95,Harlingen,Kasi00}.
Since the pair potential of the
$d_{x^{2}-y^{2}}$-wave superconductor
is anisotropic, the amplitude of the pair potential
near the surface or interface is significantly reduced.
This suppression of the pair potential causes
a very interesting phenomena, $i$.$e$.
the formation of Andreev bound state (ABS)
at the Fermi energy (zero-energy) near a
specularly reflecting  surface \cite{Hu}
when the angle between the lobe direction of the
$d_{x^{2}-y^{2}}$-wave pair potential and the normal to the interface
is nonzero.
This state is originated from
the interference effect in the effective pair potential of
the $d_{x^{2}-y^{2}}$-wave symmetry
through the reflection at the  surface or interface.
The ABS  manifests itself as a sharp peak in the middle of
the tunneling conductance spectra,
the so-called zero-bias conductance peak (ZBCP),
\cite{Tanaka95}
and the consistency between theory and experiments
has been checked in details
\cite{Kasi00,Kashiwaya95,Kashiwaya96,Tanuma1,Alff,Wei,Yeh,Wang,Iguchi}.
\par
On the other hand,
there still remains a controversial issue;
formation of a broken time reversal symmetry state
(BTRSS) at low temperature due to the mixing of
subdominant $s$-,\cite{SBL95,Kuboki,Sigrist98}
or $d_{xy}$-wave component
\cite{Krishana,Balatsky,Laughlin}
as the imaginary part of the pair potential
to the predominant $d_{x^{2}-y^{2}}$-wave component \cite{Matsumoto}.
Theoretical studies
based on the quasiclassical approximation
\cite{Matsumoto,Fogel}
and several lattice models \cite{Tanuma2,Zhu}
reported the presence of the induced subdominant pair potential
near the surface which breaks
the time reversal symmetry \cite{Matsumoto}.
The resulting surface density of states
with the BTRSS shows the splitting of
the zero-energy peak
\cite{Matsumoto}
and the corresponding tunneling conductance
shows the ZBCP splitting.
It has also been clarified that
the magnitude of the splitting depends on
the induced subdominant pair potential
near the surface.
Actually,
Covington $et$ $al.$
\cite{Covington},
Krupke $et$ $al.$ \cite{Krupke},
and Sharoni $et$ $al.$ \cite{Sharoni}
reports the ZBCP splitting
at low temperature and they ascribed
the origin of the ZBCP splitting
to the above BTRSS \cite{Kashi99}.
However, at the same time,
there are many experiments which do not show
the ZBCP splitting, and in these experiments,
ZBCP survives even at low temperatures
\cite{Kasi00,Wei,Wang,Iguchi,Yeh}.
Although there are several pre-existing theories
which discuss the BTRSS \cite{Matsumoto},
almost all these theories treat the semi-infinite
$d_{x^{2}-y^{2}}$-wave superconductor
and the relevance to the
actual experiments of tunneling spectroscopy
has not been fully clarified yet.
At this stage, it is an important problem to
clarify the stability and the possible
observability  of the formation of BTRSS,
$i.e.$, the ZBCP  splitting,
for various condition of the junctions.
There are several factors which
determine the magnitude of
the splitting of ZBCP;
(i) inter-electron potential
which induces the subdominant pair potential near the interface,
(ii) transmission probability of the particles at the interface,
(iii) orientation of the junctions,
$i.e.$, the angle between the normal to the interface
and the crystal axis of
$d_{x^{2}-y^{2}}$-wave superconductor,
(iv) finite temperature effect,
(v) roughness of the interface \cite{Rainer},
and (vi) impurity concentration
in the superconductor \cite{Tanaka01}.
\par
Although there are several factors which determine 
the magnitude of the  splitting of ZBCP, in this paper,
we concentrate on (i), (ii), (iii), and (iv).
First,
we calculate the spatial variation of the pair potential
in the normal metal / $d_{x^{2}-y^{2}}$ -wave superconductor
junction (N/D junction)
in the presence of subdominant pair potential
using a quasiclassical formalism
for various conditions of the junctions.
Using these results, we calculate tunneling conductance.
\par
The organization of this paper is as follows.
In Sec.~\ref{sec:2}, a formulation to calculate
the spatial dependence of the pair potential,
and the tunneling conductance
is presented.
In Sec.~\ref{sec:3},
results of the numerical calculations for
both $d_{x^{2}-y^{2}}$+i$s$-
and $d_{x^{2}-y^{2}}$+i$d_{xy}$-wave
states are discussed in detail.
Section~\ref{sec:4} is devoted
to conclusions and future problems.
\par
\section{Theoretical formulation}
\label{sec:2}
We study N/D junctions
in the clean limit,
where the normal metal is located at $x<0$
and the $d_{x^{2}-y^{2}}$-wave superconductor
extends elsewhere. 
For the simplicity,
two dimensional system is assumed and
the $x$-axis is taken
perpendicular to the flat interface
located at $x=0$.
When quasiparticles are in the $xy$-plane, 
a transmitted electron-like quasiparticle 
and hole-like quasiparticle feel 
different effective pair potentials
$\Delta(\phi_{+})$ and $\Delta(\phi_{-})$,
with $\phi_{+}=\phi$ and $\phi_{-}=\pi-\phi$.
Here, 
$\phi$ is the azimuthal angle
in the $xy$-plane
given by $(k_{x}+{\rm i}k_{y})/|{\vct k}|={\rm e^{i\phi}}$.
The barrier potential at the interface
has a $\delta$-functional  form $H\delta(x)$,
where $\delta(x)$ and $H$ are
the $\delta$-function and its amplitude, respectively.
A cylindrical Fermi surface is assumed 
and the magnitude of the Fermi momentum 
and the effective mass are chosen to be equal 
both in the normal metal 
and in the superconductor.
\par
The quasiclassical Green's function method 
\cite{Matsumoto,Fogel,Eilen,Bruder,Ohashi,QCM,Buch,Barash,Boss01}
developed by Ashida $et$ $al.$
\cite{Ashida,Nagato}
is used in order to determine 
the spatial variation of the pair potential
self-consistently.
In the following,
we  briefly summarize this
quasiclassical method we used.
We start with the Bogoliubov-de Gennes (BdG)
equation for unconventional 
spin-singlet superconductors
\cite{Bruder,BdG},
\begin{eqnarray}
E_{n}\tilde{u}_{n}({\vct r}) &=& H_{0}\tilde{u}_{n}({\vct r})
+ \int d{\vct r^{\prime}}
\Delta({\vct r},{\vct r^{\prime}})
\tilde{v}_{n}({\vct r^{\prime}}),
\\
E_{n}\tilde{v}_{n}({\vct r}) &=& -H_{0}\tilde{v}_{n}({\vct r})
+ \int d{\vct r^{\prime}}
\Delta^{*}({\vct r},{\vct r^{\prime}})
\tilde{u}_{n}({\vct r^{\prime}}), \\
H_{0} &=& -\frac{\hbar^{2}}{2m}\nabla^{2} - \mu,
\end{eqnarray}
where $\mu$ is the chemical potential,
while $\tilde{u}_{n}({\vct r})$ and $\tilde{v}_{n}({\vct r})$
denote the electron-like and hole-like components
of the wave function,
\begin{eqnarray}
\tilde{\Psi}_{n}({\vct r}) &=&
\left (
    \begin{array}{c}
        \tilde{u}_{n}({\vct r}) \\
        \tilde{v}_{n}({\vct r})
    \end{array}
\right ), \nonumber \\
&\equiv &
\left (
    \begin{array}{c}
        u_{n}(\hat{\vct k},{\vct r}) \\
        v_{n}(\hat{\vct k},{\vct r})
    \end{array}
\right ) e^{{\rm i}k_{\rm F} \cdot r}
=\Psi_{n}(\hat{\vct k},{\vct r})
e^{{\rm i}k_{\rm F} \cdot r}.
\end{eqnarray}
Here the quantities
$\hat{\vct k}$ and ${\vct r}$
stand for the unit vector of
the wave number of the Cooper pair which
is fixed on the Fermi surface
$(\hat{\vct k}={\vct k}_{\rm F}/|{\vct k}_{\rm F}|)$,
and the  position of the center of mass of Cooper pair,
respectively. 
After applying the quasiclassical approximation,
the BdG equation is reduced to the Andreev equation
\cite{Hu,Bruder,Kurki},
\begin{eqnarray}
E_{n}\Psi_{n}(\hat{\vct k},{\vct r})
   &=&-\left [ {\rm i}\hbar v_{\rm F}{\vct k \cdot \nabla}
      + \hat{\Delta}(\hat{\vct k},{\vct r}) \right ]
       \hat{\tau}_{3}\Psi_{n}(\hat{\vct k},{\vct r}), \\
\hat{\Delta}(\hat{\vct k},{\vct r}) &=&
\left (
    \begin{array}{cc}
        0 & \Delta(\hat{\vct k},{\vct r}) \\
  -\Delta^{*}(\hat{\vct k},{\vct r}) & 0
    \end{array}
\right ),
\end{eqnarray}
where $v_{\rm F}$ and $\hat{\tau}_{i}(i=1,2,3)$
stand for Fermi velocity and Pauli matrices, respectively.
The wave function $\Psi_{n}(\hat{\vct k},{\vct r})$
is obtained by neglecting the rapidly oscillating
plane-wave part following the quasiclassical approximation
\cite{Bruder,Kurki}.
The $\hat{\vct k}$ dependence of
$\Delta(\hat{\vct k},{\vct r})$
represents the symmetry of the pair potential.
\par
Now, we consider the case where a specularly
reflecting surface or interface runs along the $y$-direction.
In this case,
the pair potential depends only on $x$
since the system is homogeneous along the $y$-direction.
It is convenient to introduce the following directional
notation,\cite{Matsumoto,Ashida}
\begin{eqnarray}
\Psi_{n}(\hat{\vct k},{\vct r}) &=&
 \Phi_{n}^{(+)}(\phi_{+},x)e^{{\rm i}|k_{{\rm F}x}|x}
+\Phi_{n}^{(-)}(\phi_{-},x)e^{-{\rm i}|k_{{\rm F}x}|x},
\nonumber \\
\Phi_{n}^{(\alpha)}(\phi_{\alpha},x) &=&
\left (
    \begin{array}{c}
        u_{n}^{(\alpha)}(\phi_{\alpha},x) \\
        v_{n}^{(\alpha)}(\phi_{\alpha},x)
    \end{array}
\right ).
\end{eqnarray}
Here $\pm$ represents
the sign of the $x$ component of the Fermi wave number
$k_{{\rm F}x}$ and $\alpha (\beta) = \pm$.
We define a Green's function
$G_{\alpha \beta}(\phi,x,x^{\prime})$
and a quasiclassical Green's function
$g_{\alpha \beta}(\phi,x)$,
\begin{eqnarray}
G_{\alpha \beta}(\phi,x,x^{\prime})
&=& \sum_{n}\frac{\Phi_{n}^{(\alpha)}(\phi_{\alpha},x)
\Phi_{n}^{(\beta)\dagger}(\phi_{\beta},x)}
{{\rm i}\omega_{m} -E_{n}}, \\
g_{\alpha \beta}(\phi,x) \pm {\rm i}
(\hat{\gamma}_{3})_{\alpha \beta}
&=& -2\hbar|v_{{\rm F}x}|\hat{\tau}_{3}
G_{\alpha \beta}(\phi,x\pm 0,x).
\end{eqnarray}
In the above,
$\hat{\gamma}_{3}$ is the Pauli matrix
in the directional space\cite{Ashida}
and $v_{{\rm F}x}$ is
the $x$-component of the Fermi velocity.
The quasiclassical Green's function
$g_{\alpha \beta}(\phi,x)$ obeys
the Eilenberger equation,\cite{Eilen}
\begin{eqnarray}
{\rm i}|v_{{\rm F}x}|\frac{\partial}{\partial x}
g_{\alpha \beta}(\phi,x) = &-&\alpha
\left [ {\rm i}\omega_{m}\hat{\tau}_{3}
+\hat{\Delta}(\phi_{\alpha},x) \right ] 
g_{\alpha \beta}(\phi,x)
\nonumber \\
+ &\beta & g_{\alpha \beta}(\phi,x)
\left [ {\rm i}\omega_{m}\hat{\tau}_{3}
+\hat{\Delta}(\phi_{\beta},x) \right ] , \\
\hat{\Delta}(\phi_{\alpha},x) &=&
\left (
    \begin{array}{cc}
        0 & \Delta(\phi_{\alpha},x) \\
  -\Delta^{*}(\phi_{\alpha},x) & 0
    \end{array}
\right ),
\end{eqnarray}
where $\omega_{m}$
is the Matsubara frequency.
The quasiclassical Green's function
can be written by the following evolution operator
$U_{\alpha}(\phi_{\alpha},x,x^{\prime})$ as
\begin{eqnarray}
g_{\alpha \beta}(\phi,x)
   =U_{\alpha}(\phi_{\alpha},x,x^{\prime})
     g_{\alpha \beta}(\phi,x^{\prime})
      U_{\beta}^{-1}(\phi_{\beta},x,x^{\prime}),
\end{eqnarray}
where $U_{\alpha}(\phi_{\alpha},x,x^{\prime})$
satisfies the Andreev equation
\begin{eqnarray}
{\rm i}\hbar |v_{{\rm F}x}|
\frac{\partial}{\partial x}
U_{\alpha}(\phi_{\alpha},x,x^{\prime})
=-\alpha
\left [ {\rm i}\omega_{m}\hat{\tau}_{3}
+\hat{\Delta}(\phi_{\alpha},x)\right ]
U_{\alpha}(\phi_{\alpha},x,x^{\prime}),
\end{eqnarray}
with $U_{\alpha}(\phi_{\alpha},x,x)=1$.
\par
Considering a semi-infinite N/D junction geometry,
the pair potential
in the superconductor side
approaches to the bulk value
$\Delta(\phi_{\alpha},\infty)$
at sufficiently large $x$.
Hence,
the evolution operator can be divided
into a growing part and a decaying part:
\begin{eqnarray}
U_{\alpha}(\phi_{\alpha},x,x^{\prime}) &=&
\Lambda_{\alpha}^{(+)}(\phi_{\alpha},x,x^{\prime})
e^{{\kappa_{\alpha}}(x-x^{\prime})}
+ \Lambda_{\alpha}^{(-)}(\phi_{\alpha},x,x^{\prime})
e^{-{\kappa_{\alpha}}(x-x^{\prime})},
\\
\Lambda_{\alpha}^{(+)}(\phi_{\alpha},x,x^{\prime})
&=& -\frac{1}{W_{\alpha}}
\Phi_{n}^{(+)}(\phi_{\alpha},x)
{}^{\rm T}\Phi_{n}^{(-)}(\phi_{\alpha},x^{\prime})
\hat{\tau}_{2},
\nonumber \\
\Lambda_{\alpha}^{(-)}(\phi_{\alpha},x,x^{\prime})
&=& \frac{1}{W_{\alpha}}
\Phi_{n}^{(-)}(\phi_{\alpha},x)
{}^{\rm T}\Phi_{n}^{(+)}(\phi_{\alpha},x^{\prime})
\hat{\tau}_{2},
\end{eqnarray}
where
\begin{eqnarray}
\kappa_{\alpha} &=& \frac{\Omega_{\alpha}}{|v_{{\rm F}x}|},
\quad
\Omega_{\alpha}=
\sqrt{\omega_{m}^{2}+|\Delta(\phi_{\alpha},\infty)|^{2}},
\nonumber \\
W_{\alpha} &=&
{}^{\rm T}\Phi_{n}^{(+)}(\phi_{\alpha},x)\hat{\tau}_{2}
\Phi_{n}^{(-)}(\phi_{\alpha},x)
=-{}^{\rm T}\Phi_{n}^{(-)}(\phi_{\alpha},x)\hat{\tau}_{2}
\Phi_{n}^{(+)}(\phi_{\alpha},x)
\nonumber \\
&=& \mbox{constant}.
\end{eqnarray}
In the above,  $^{\rm T}\Phi_{n}^{(\alpha)}(\phi_{\alpha},x)$
denotes  the transposition of $\Phi_{n}^{(\alpha)}(\phi_{\alpha},x)$.
\par
Retaining the most divergent term
in semi-infinite limit,
we find
the quasiclassical Green's function
$\hat{g}_{\alpha \alpha}(\phi,x)$
in the superconductor side
given by \cite{QCM,Nagato},
\begin{eqnarray}
\hat{g}_{\alpha \alpha}(\phi_{\alpha},x)
={\rm i}
\left (
\frac{2\hat{A}_{{\rm S}\alpha}(x)}
{{\rm Tr}[\hat{A}_{{\rm S}\alpha}(x)]}
-1
\right ),
\end{eqnarray}
where
\begin{eqnarray}
\label{as+}
\hat{A}_{{\rm S}+}(x) &=&
\Lambda_{+}^{(-)}(\phi_{+},x,L)
\Lambda_{-}^{(+)}(\phi_{-},L,0)
\hat{R}_{\rm N}
U_{+}(\phi_{+},0,x)e^{-\kappa x}
\nonumber \\
&=&\hat{\lambda}_{{\rm S}+}(x,0)\hat{R}_{\rm N}
\tilde{U}_{+}(\phi_{+},0,x),
\\
\hat{\lambda}_{{\rm S}+}(x,0) & \propto &
\left (
\begin{array}{c}
      u_{n}^{(-)}(\phi_{+},x) \\
      v_{n}^{(-)}(\phi_{+},x) \\
\end{array}
\right )
\left (
\begin{array}{cc}
      u_{n}^{(-)}(\phi_{-},0) &
      v_{n}^{(-)}(\phi_{-},0) \\
\end{array}
\right )
\hat{\tau_{2}}.
\end{eqnarray}
In the above,
the matrix $\hat{R}_{\rm N}$ represents
resistance at the interface which is given by
\cite{Boss01}
\begin{eqnarray}
\hat{R}_{\rm N} \propto
\left (
\begin{array}{cc}
1 & 0 \\
0 & 1-\sigma_{\rm N}(\phi)
\end{array}
\right ),
\quad
\sigma_{\rm N}(\phi)=
\frac{4\cos \phi^{2}}{4\cos \phi^{2} + Z^{2}},
\end{eqnarray}
where $\sigma_{\rm N}(\phi)$
stands for tunneling conductance
when the system is in the normal state
\cite{Boss02}
and $Z$ is the effective barrier height
at the interface with $Z=2mH/(\hbar^{2}k_{\rm F})$.
In order to obtain the quantity
$\tilde{U}_{+}(\phi_{+},0,x)$
in eq.(\ref{as+}),
we rewrite $\hat{A}_{{\rm S}+}(x)$ as
\cite{Nagato}
\begin{eqnarray}
\hat{A}_{{\rm S}+}(x) =
\left (
\begin{array}{c}
      u_{n}^{(-)}(\phi_{+},x) \\
      v_{n}^{(-)}(\phi_{+},x) \\
\end{array}
\right )
\left (
\begin{array}{cc}
      X_{+}(x) &
      Y_{+}(x) \\
\end{array}
\right )
\hat{\tau_{2}},
\end{eqnarray}
where
${\cal D}_{\alpha}(x)=
(-{\rm i})v_{n}^{(-)}(\phi_{\alpha},x)
/u_{n}^{(-)}(\phi_{\alpha},x)$
and
${\cal F}_{+}(x)={\rm i}X_{+}(x)/Y_{+}(x)$
obey the following Riccati type equations
\begin{eqnarray}
\label{req}
\hbar |v_{{\rm F}x}|
\frac{\partial}{\partial x}{\cal D}_{\alpha}(x)
&=& \alpha \left [
-2\omega_{m} {\cal D}_{\alpha}(x) +
\Delta(\phi_{\alpha},x){\cal D}_{\alpha}^{2}(x)
- \Delta^{*}(\phi_{\alpha},x)
\right ],
\\
\hbar |v_{{\rm F}x}|
\frac{\partial}{\partial x}{\cal F}_{+}(x)
&=&
2\omega_{m} {\cal F}_{+}(x) +
\Delta^{*}(\phi_{+},x){\cal F}_{+}^{2}(x)
- \Delta(\phi_{+},x).
\end{eqnarray}
We can write the quasiclassical Green's function
in a compact form \cite{QCM},
\begin{eqnarray}
\label{g++}
\hat{g}_{++}(\phi_{+},x) &= {\rm i}
\left [ {\displaystyle
\frac{2}{1-{\cal D}_{+}(x){\cal F}_{+}(x)}}
\left (
\begin{array}{cc}
1 & {\rm i}{\cal F}_{+}(x) \\
{\rm i}{\cal D}_{+}(x) & -{\cal D}_{+}(x){\cal F}_{+}(x)
\\
\end{array}
\right )
-1
\right ].
\end{eqnarray}
Initial conditions of these equations are
\begin{eqnarray}
{\cal D}_{\alpha}(\infty)=
\frac{\Delta^{*}(\phi_{\alpha},\infty)}
{\omega_{m} + \alpha\Omega_{\alpha}},
\quad
{\cal F}_{+}(0)=\frac{1-\sigma_{\rm N}(\phi)}
{D_{-}(0)}.
\end{eqnarray}
\par
The pair potential is given by
\cite{Matsumoto,Ashida,Nagato,Boss01,Bruder,Ohashi}
\begin{eqnarray}
\label{peq}
\Delta(\phi,x) =
\sum_{0 \leq m < \omega_{c}/2\pi T}
\frac{1}{2\pi}
{\displaystyle
\int ^{\pi/2}_{-\pi/2}
d\phi^{\prime}\sum_{\alpha}V(\phi,\phi^{\prime}_{\alpha})}
[\hat{g}_{\alpha \alpha}(\phi^{\prime}_{\alpha},x)]_{12},
\end{eqnarray}
with 
$\hat{g}_{--}(\phi_{-},x) = -\hat{g}_{++}(-\phi_{+},x)$, 
where $\omega_{c}$ is the cutoff energy and
$[\hat{g}_{\alpha \alpha}(\phi_{\alpha},x)]_{12}$
means the 12 element of
$\hat{g}_{\alpha \alpha}(\phi_{\alpha},x)$.
Here $V(\phi,\phi_{\alpha})$
is the effective inter-electron
potential of the Cooper pair.
In our numerical calculations,
new $\Delta(\phi_{\alpha},x)$ and 
$\hat{g}_{\alpha \alpha}(\phi_{\alpha},x)$ are obtained
using eqs.(\ref{req})-(\ref{g++}) and eq. (\ref{peq}).
We repeat this iteration process
until the sufficient convergence is obtained.
\par
Next, we calculate
the tunneling conductance spectra based on the self-consistently
determined pair potential.
The resulting  normalized tunneling conductance
$\sigma_{\rm T}(eV)$ with the bias voltage $V$
is given by \cite{Tanaka01,Boss02}
\begin{eqnarray}
\sigma_{\rm T}(eV) &=&
\frac{\displaystyle
\int^{\pi/2}_{-\pi/2}d\phi
\int^{\infty}_{-\infty}dE
\sigma_{\rm N}(\phi)
\sigma_{\rm S}(E,\phi)
{\rm sech}^{2}\left ( \frac{E-eV}{2k_{\rm B}T} \right )
\cos \phi}
{\displaystyle
\int^{\pi/2}_{-\pi/2}d\phi 
\int^{\infty}_{-\infty} dE 
\sigma_{\rm N}(\phi)
{\rm sech}^{2}\left ( \frac{E-eV}{2k_{\rm B}T} \right )
\cos \phi},
\\
\sigma_{\rm S}(E,\phi)
&=&
\frac{1+\sigma_{\rm N}(\phi)|\Gamma_{\rm S+}(E,\phi_{+},0)|^{2}
+[\sigma_{\rm N}(\phi)-1]
|\Gamma_{\rm S+}(E,\phi_{+},0)|^{2}|\Gamma_{\rm S-}(E,\phi_{-},0)|^{2}}
{|1+[\sigma_{\rm N}(\phi)-1]
\Gamma_{\rm S+}(E,\phi_{+},0)\Gamma_{\rm S-}(E,\phi_{-},0)|^{2}}.
\end{eqnarray}
In the above,
$\Gamma_{{\rm S}\alpha}(E,\phi_{\alpha},x)$
is obtained by solving following equations,
\begin{eqnarray}
{\rm i}\hbar |v_{{\rm F}x}|
\frac{\partial}{\partial x}\Gamma_{\rm S+}(E,\phi_{+},x)
=2E \Gamma_{\rm S+}(E,\phi_{+},x) -
\Delta(\phi_{+},x)\Gamma_{\rm S+}^{2}(E,\phi_{+},x)
-\Delta^{*}(\phi_{+},x),
\\
{\rm i}\hbar |v_{{\rm F}x}|
\frac{\partial}{\partial x}\Gamma_{\rm S-}(E,\phi_{-},x)
=2E \Gamma_{\rm S-}(E,\phi_{-},x) -
\Delta^{*}(\phi_{-},x)\Gamma_{\rm S-}^{2}(E,\phi_{-},x)
-\Delta(\phi_{-},x).
\end{eqnarray}
In the following, 
almost calculations are performed 
on the temperature  $T/T_{c}=0.05$,
where $T_{c}$ is the critical temperature of
the bulk $d_{x^{2}-y^{2}}$-wave superconductor.
\par
\section{Broken time reversal symmetry state
near an interface of
a $d_{x^{2}-y^{2}}$-wave superconductor}
\label{sec:3}
In this section,
the spatial dependence of the
self-consistently determined pair potential
and the corresponding tunneling conductance
are presented for
the $d_{x^{2}-y^{2}}$-wave superconducting state.
In the middle of the 
$d_{x^{2}-y^{2}}$-wave superconductor,
the pair potential is given by
$\Delta(\phi_{\alpha},\infty)=\Delta_{0}\cos [2(\phi-\theta)]$,
where $\theta$ is the 
angle between normal to the interface and
the lobe direction of the $d_{x^{2}-y^{2}}$-wave pair potential,
$i.e.$, the angle between 
the $x$-axis and the crystal $a$-axis of
the $d_{x^{2}-y^{2}}$-wave.
In this paper, 
we choose various $\theta$ $(0 \leq \theta \leq \pi/4)$ 
by changing the magnitude of 
$Z$ and $T_{s}$ ($T_{d_{xy}}$). 
\par
%
\subsection{$d_{x^{2}-y^{2}}${\rm + i}$s$-wave state}
\begin{figure}[t]
\begin{center}
\begin{minipage}[t]{8cm}
\epsfxsize=8cm
\epsfbox{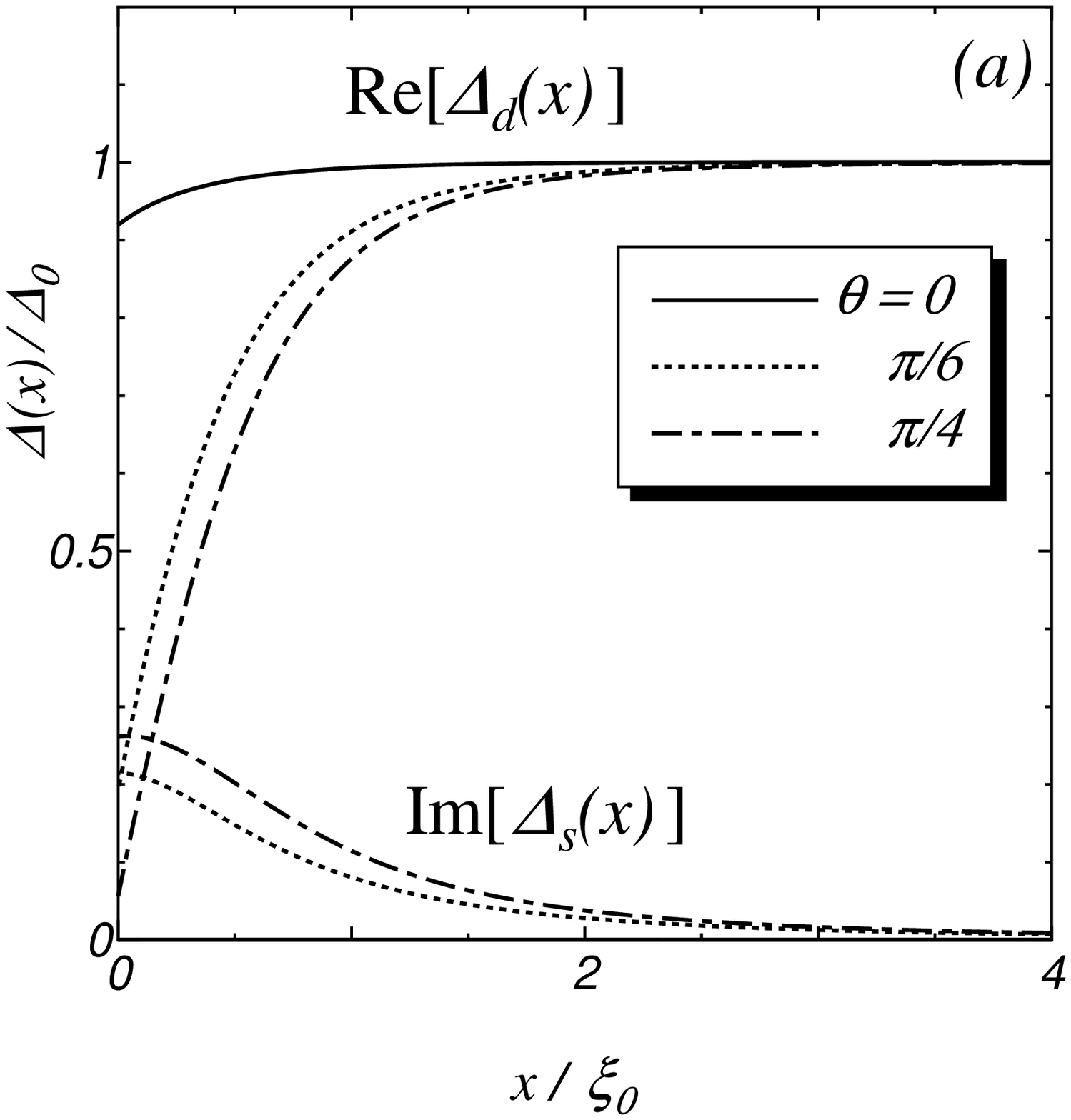}
\end{minipage}
\begin{minipage}[t]{8cm}
\epsfxsize=8cm
\epsfbox{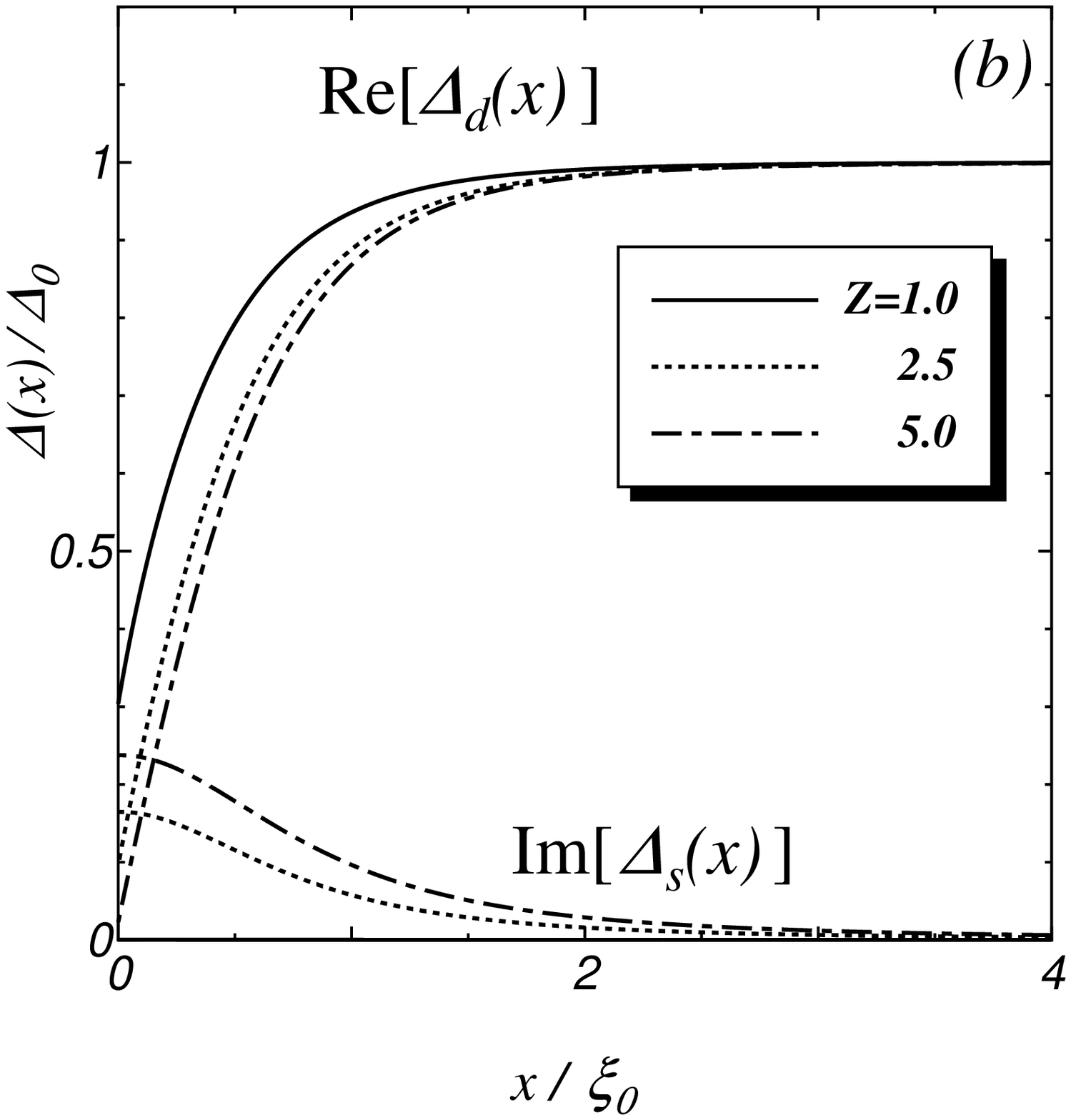}
\end{minipage}
\end{center}
\caption{Spatial dependences of the pair potential
at $T_{s}/T_{d}=0.2$;
(a) near the interface for the angle $\theta$
between $x$-axis and crystal axis at $Z=3.0$,
and
(b) near the (110) interface [$\theta=\pi/4$] 
for various height of barrier $Z$. $T=0.05T_{c}$.
\label{fig01}}
\end{figure}

In this subsection,
we show the spatial dependence of the pair potential
and the resulting tunneling conductance of
the $d_{x^{2}-y^{2}}$+i$s$-wave state realized
near the interface of the N/D junction.
The spatial dependence of the 
pair potential is expressed as
\begin{eqnarray}
\Delta(\phi,x)=\Delta_{d}(x)\cos [2(\phi-\theta)] 
+\Delta_{s}(x),
\end{eqnarray}
where $\Delta_{d}(x)$ and $\Delta_{s}(x)$
correspond to the amplitude of the $d_{x^{2}-y^{2}}$-wave and
$s$-wave superconducting states, respectively.
The attractive potential $V(\phi,\phi^{\prime})$ is 
given by
\begin{eqnarray}
V(\phi,\phi^{\prime}) = 
2V_{d}\cos[2(\phi-\theta)]\cos[2(\phi^{\prime}-\theta)] 
+ V_{s}
\end{eqnarray}
where $V_{d}$ and $V_{s}$ denote the attractive 
potential of predominant $d_{x^{2}-y^{2}}$-wave 
and subdominant $s$-wave, respectively,
and they are given as
\begin{eqnarray}
V_{d} &=& \frac{2\pi k_{\rm B}T}
{\displaystyle \log \frac{T}{T_{c}}
+ \sum_{0 \leq m < \omega_{c}/2\pi T}\frac{1}{m+1/2} },
\\
V_{s} &=& \frac{2\pi k_{\rm B}T}
{\displaystyle \log \frac{T}{T_{s}}
+ \sum_{0 \leq m < \omega_{c}/2\pi T}\frac{1}{m+1/2} }.
\end{eqnarray}
Here, $T_{s}$ denotes the transition temperature of 
$s$-wave component of the pair potential without 
predominant $d_{x^{2}-y^{2}}$-wave component.  
For $\theta=0$ or $\theta=\pi/4$, 
only $\mbox{Re}[\Delta_{d}(x)]$ and 
$\mbox{Im}[\Delta_{s}(x)]$ are nonzero.  
The spatial dependence of the pair potentials 
$\mbox{Re}[\Delta_{d}(x)]$ and 
$\mbox{Im}[\Delta_{s}(x)]$ is plotted in 
Fig.~\ref{fig01}(a)
for various $\theta$ with $T_{s}/T_{d}=0.2$ 
and $Z=3$.
The $x$-axis of Fig.~\ref{fig01}(a)
is normalized by $\xi_{0}=\hbar v_{\rm F}/\Delta_{0}$
which is the coherence length of the superconductor.
For $\theta =0$ [(100) surface],
since $\mbox{Im}[\Delta_{s}(x)]=0$ is satisfied, 
the time reversal symmetry is not broken 
and the amplitude of $\mbox{Re}[\Delta_{d}(x)]$
is not suppressed at the interface. 
By changing the angle $\theta$ from zero,
the magnitude of $\mbox{Re}[\Delta_{d}(x)]$
is reduced near the interface,
while $\mbox{Im}[\Delta_{s}(x)]$ is
induced at the interface \cite{Matsumoto}.
The suppression of $\mbox{Re}[\Delta_{d}(x)]$
is originated from a depairing effect that the effective
pair potentials $\Delta(\phi_{+},0)$
and $\Delta(\phi_{-},0)$
have reversed contribution
to the pairing interaction for
certain range of $\phi$ for $\theta \neq 0$.
When $\mbox{Re}[\Delta_{d}(x)]$ is suppressed at the interface,
the quasiparticle forms the ABS near the interface at zero-energy \cite{Hu}. 
The ABS is unstable with the introduction of $s$-wave attractive potential,
then $\mbox{Im}[\Delta_{s}(x)]$ is induced at the interface \cite{Matsumoto}.
The magnitude of $\mbox{Im}[\Delta_{s}(x)]$
becomes maximum at $\theta =\pi/4$,
where the above suppression effect is most significant.
In Fig.~\ref{fig01}(b),
$\mbox{Re}[\Delta_{d}(x)]$ and 
$\mbox{Im}[\Delta_{s}(x)]$ are plotted
for various $Z$ with $T_{s}/T_{d}=0.2$
and $\theta=\pi/4$.
Even if the BTRSS becomes to be most stable
at $\theta =\pi/4$,
when the height of barrier is small,
the magnitude of the subdominant
imaginary component of $\Delta_{s}(x)$
is not induced at all.
The induced imaginary component of $\Delta_{s}(x)$
is enhanced with the increase of $Z$.
\par
\begin{figure}[t]
\begin{center}
\begin{minipage}[t]{8cm}
\epsfxsize=8cm
\epsfbox{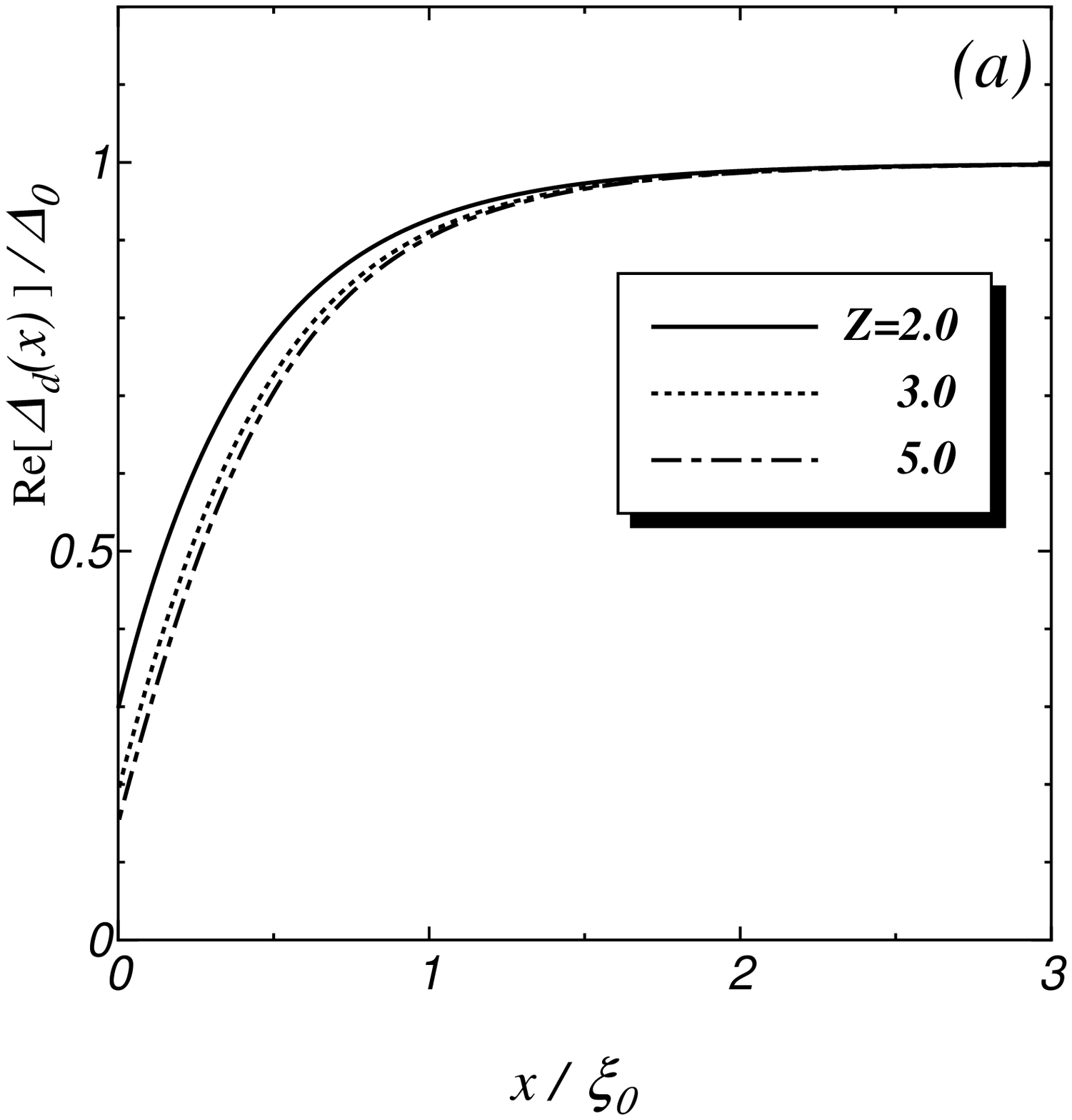}
\end{minipage}
\begin{minipage}[t]{8cm}
\epsfxsize=8cm
\epsfbox{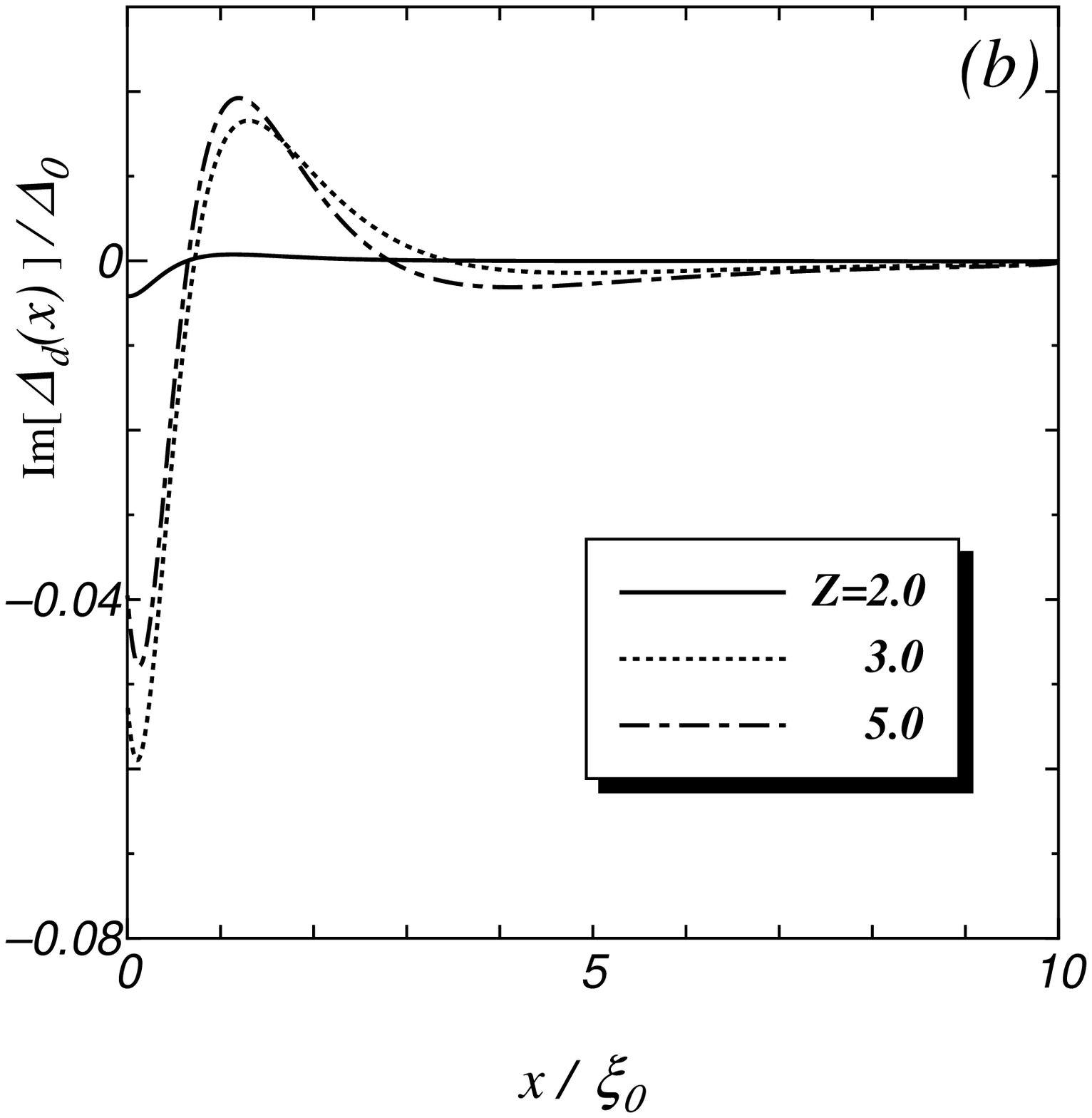}
\end{minipage}
\begin{minipage}[t]{8cm}
\epsfxsize=8cm
\epsfbox{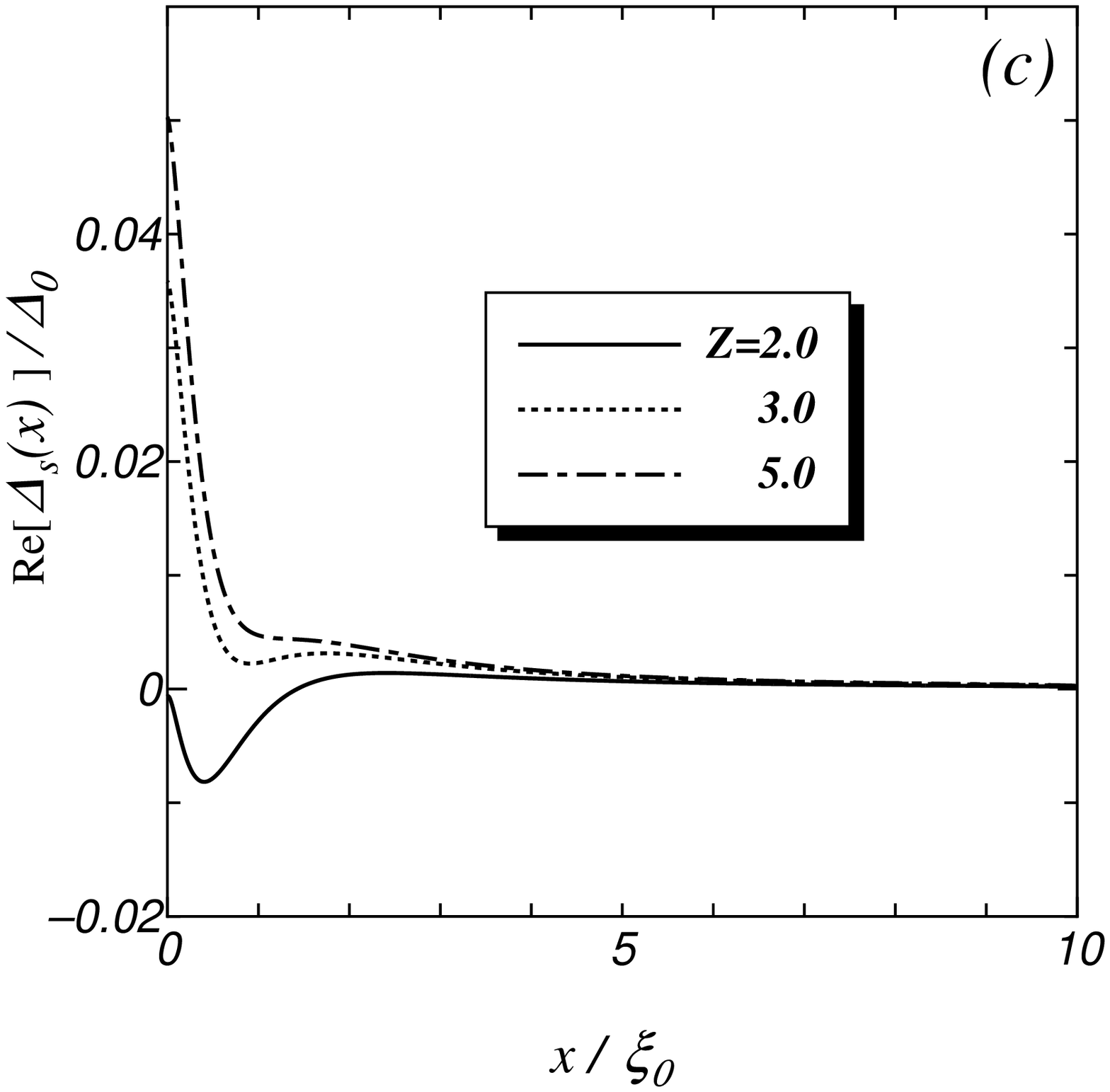}
\end{minipage}
\begin{minipage}[t]{8cm}
\epsfxsize=8cm
\epsfbox{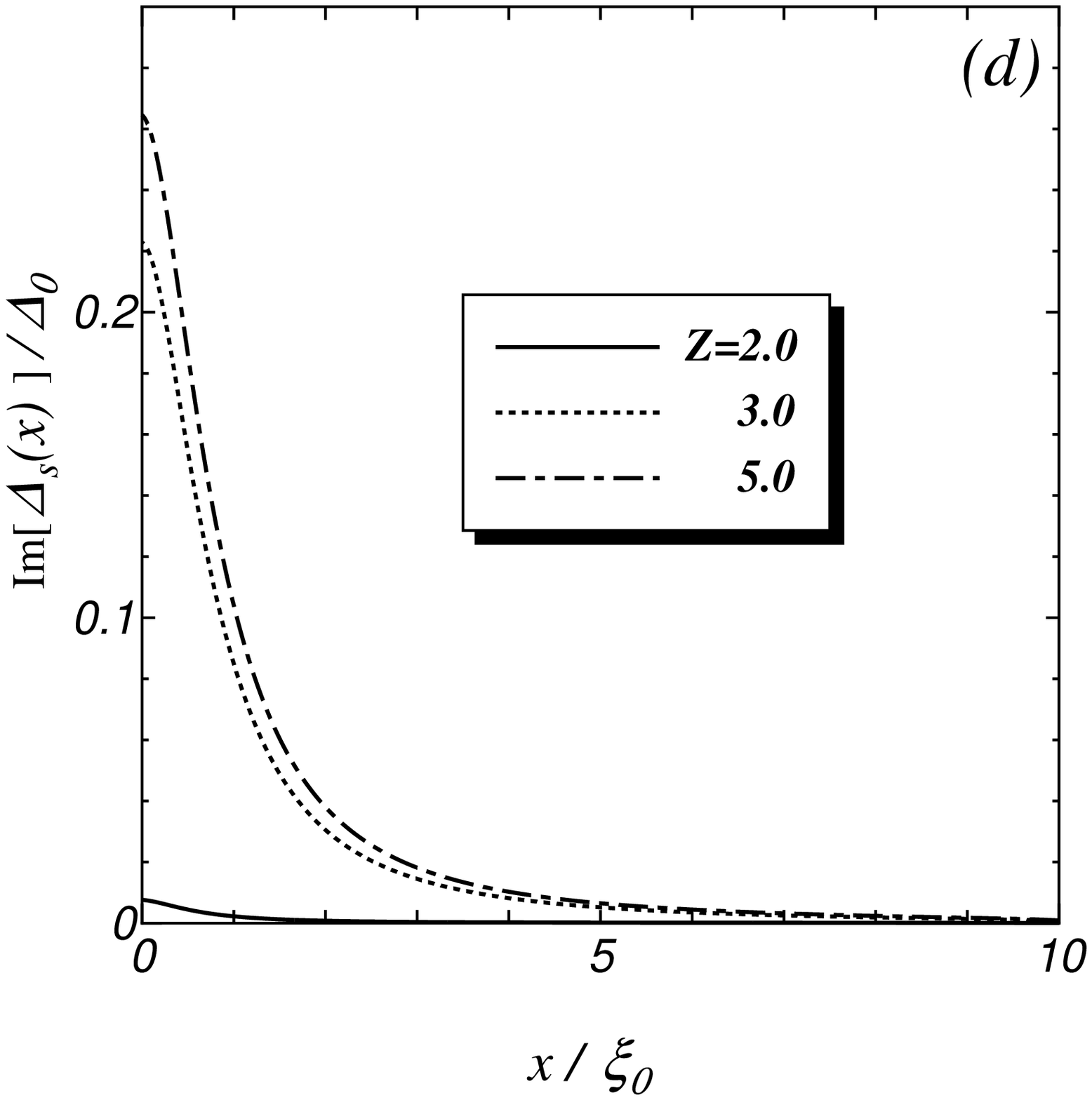}
\end{minipage}
\end{center}
\caption{
Spatial dependence of the pair potentials
near the interface with $\theta =\pi/6$
and $T_{s}/T_{d}=0.3$;
(a) real part of $\Delta_{d}(x)$,
(b) imaginary part of $\Delta_{d}(x)$,
(c) real part of $\Delta_{s}(x)$,
and
(d) imaginary part of $\Delta_{s}(x)$. 
$T=0.02T_{c}$.
\label{fig02}}
\end{figure}

The spatial dependence of the pair potentials
near the interface with the intermediate angle ($\theta =\pi/6$)
is shown in Fig.~\ref{fig02} for various height of barrier. 
In such a case, 
both $\mbox{Im}[\Delta_{d}(x)]$ and 
$\mbox{Re}[\Delta_{s}(x)]$ becomes nonzero and  
the spatial dependence is much more complex as 
compared to that for $\theta=0$ or $\theta=\pi/4$. 
The amplitudes of
$\mbox{Im}[\Delta_{d}(x)]$, 
$\mbox{Re}[\Delta_{s}(x)]$, and $\mbox{Im}[\Delta_{s}(x)]$ 
are enhanced for larger magnitude of $Z$, 
where the suppression of $\mbox{Re}[\Delta_{d}(x)]$ is significant.  
However, the amplitudes of
$\mbox{Im}[\Delta_{d}(x)]$ and 
$\mbox{Re}[\Delta_{s}(x)]$
are one order smaller than that of $\mbox{Im}[\Delta_{s}(x)]$.  
\par
\begin{figure}[t]
\begin{center}
\begin{minipage}[t]{8cm}
\epsfxsize=8cm
\epsfbox{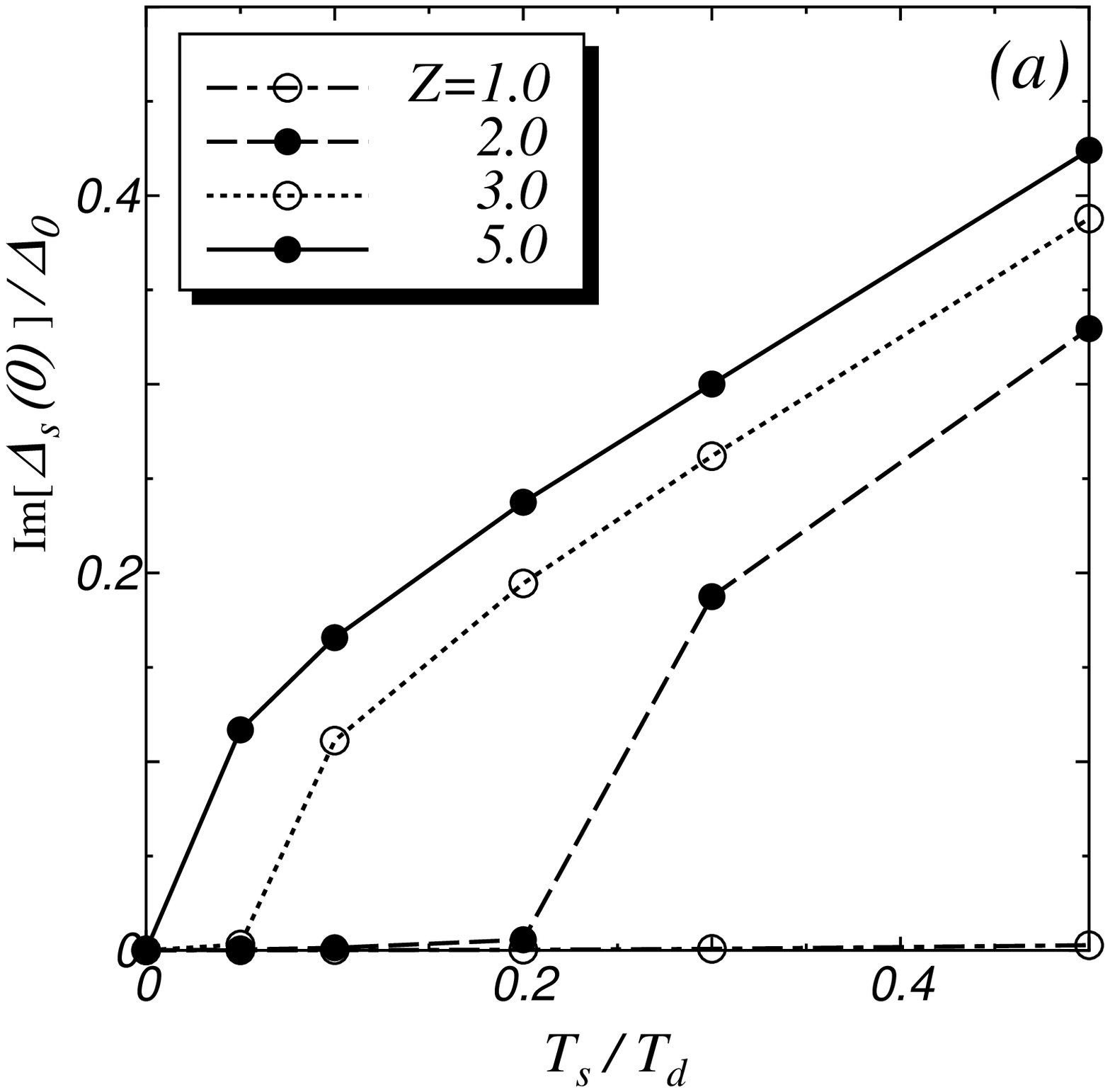}
\end{minipage}
\begin{minipage}[t]{8cm}
\epsfxsize=8cm
\epsfbox{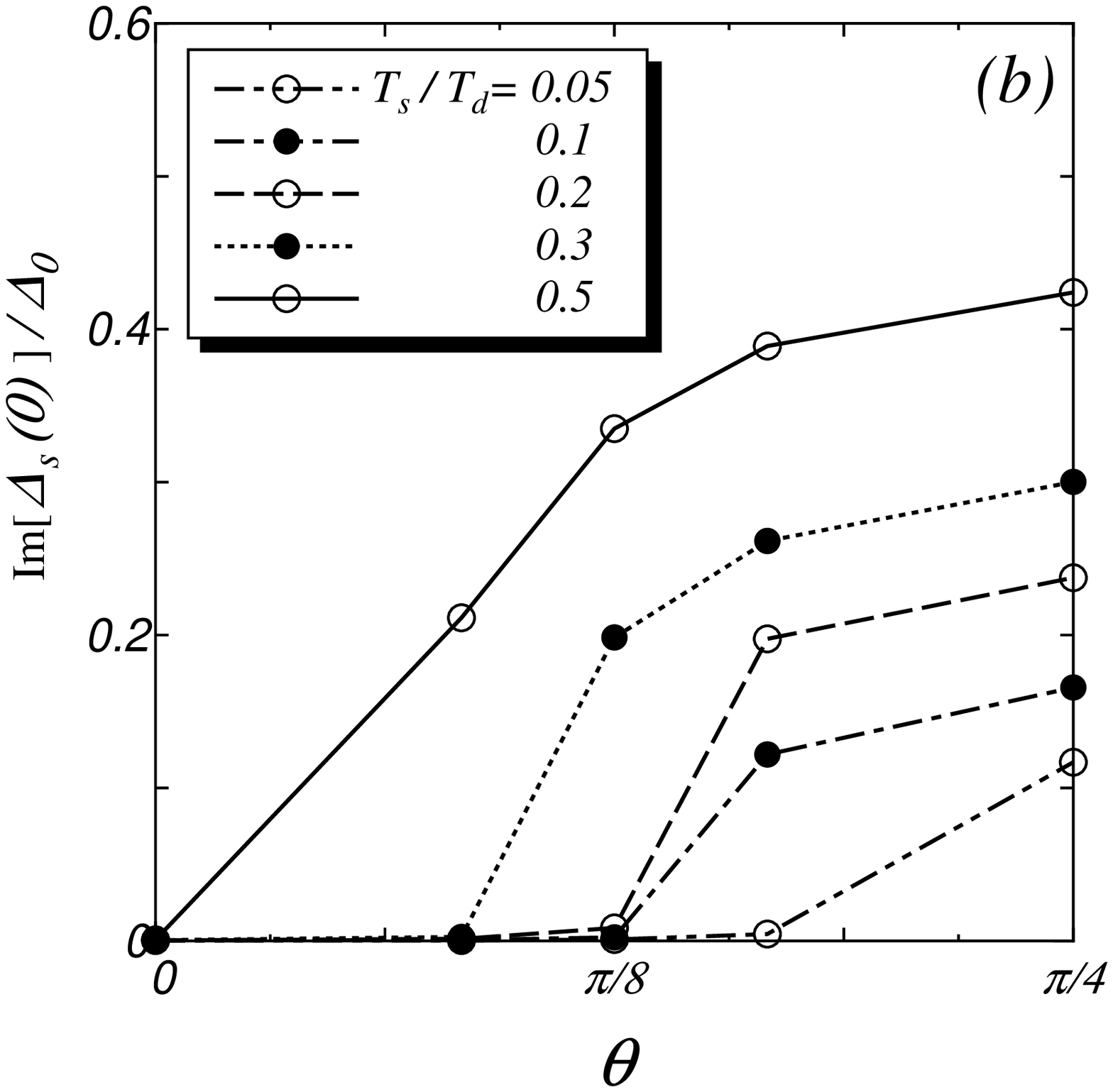}
\end{minipage}
\begin{minipage}[t]{8cm}
\epsfxsize=8cm
\epsfbox{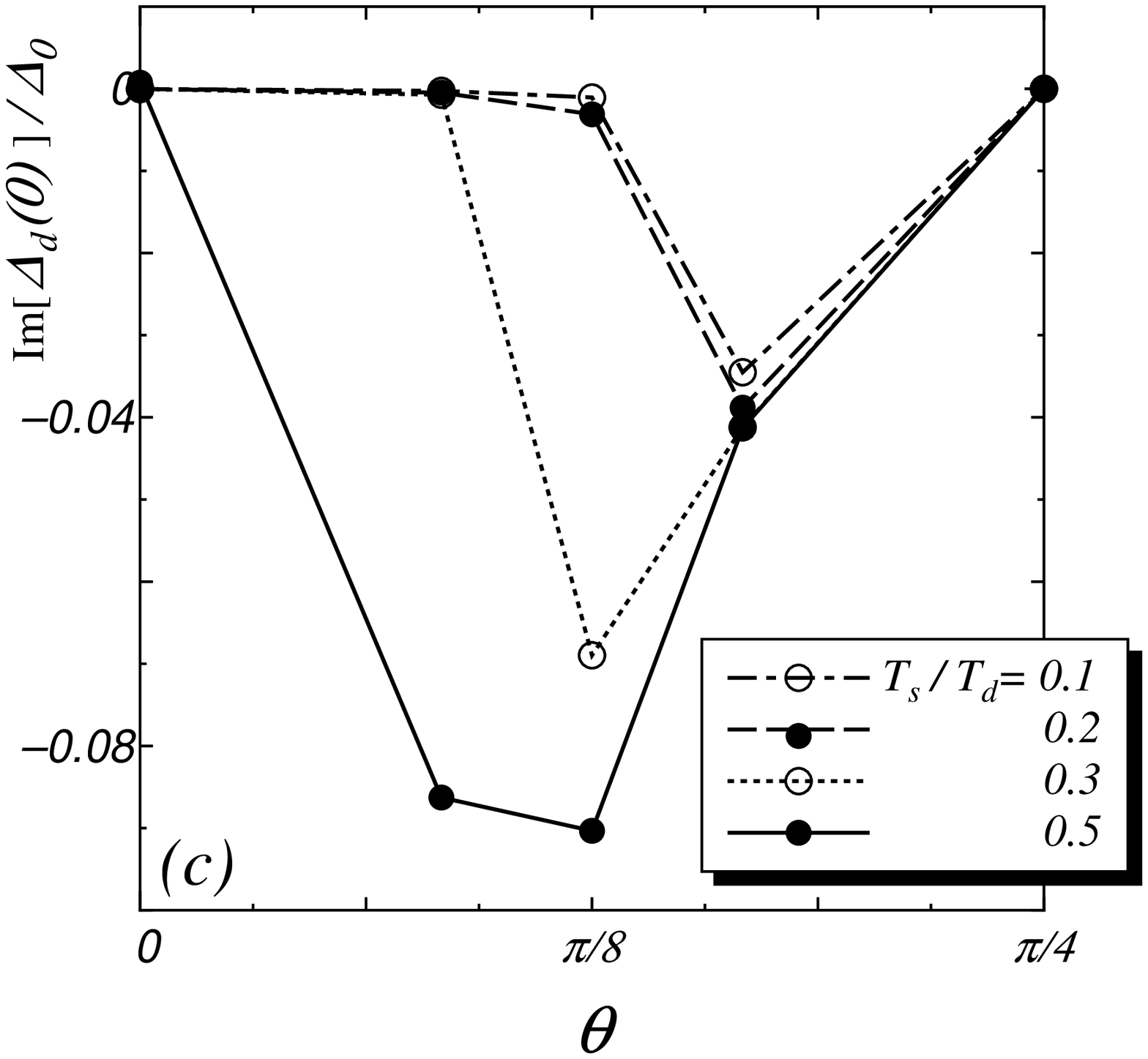}
\end{minipage}
\begin{minipage}[t]{8cm}
\epsfxsize=8cm
\epsfbox{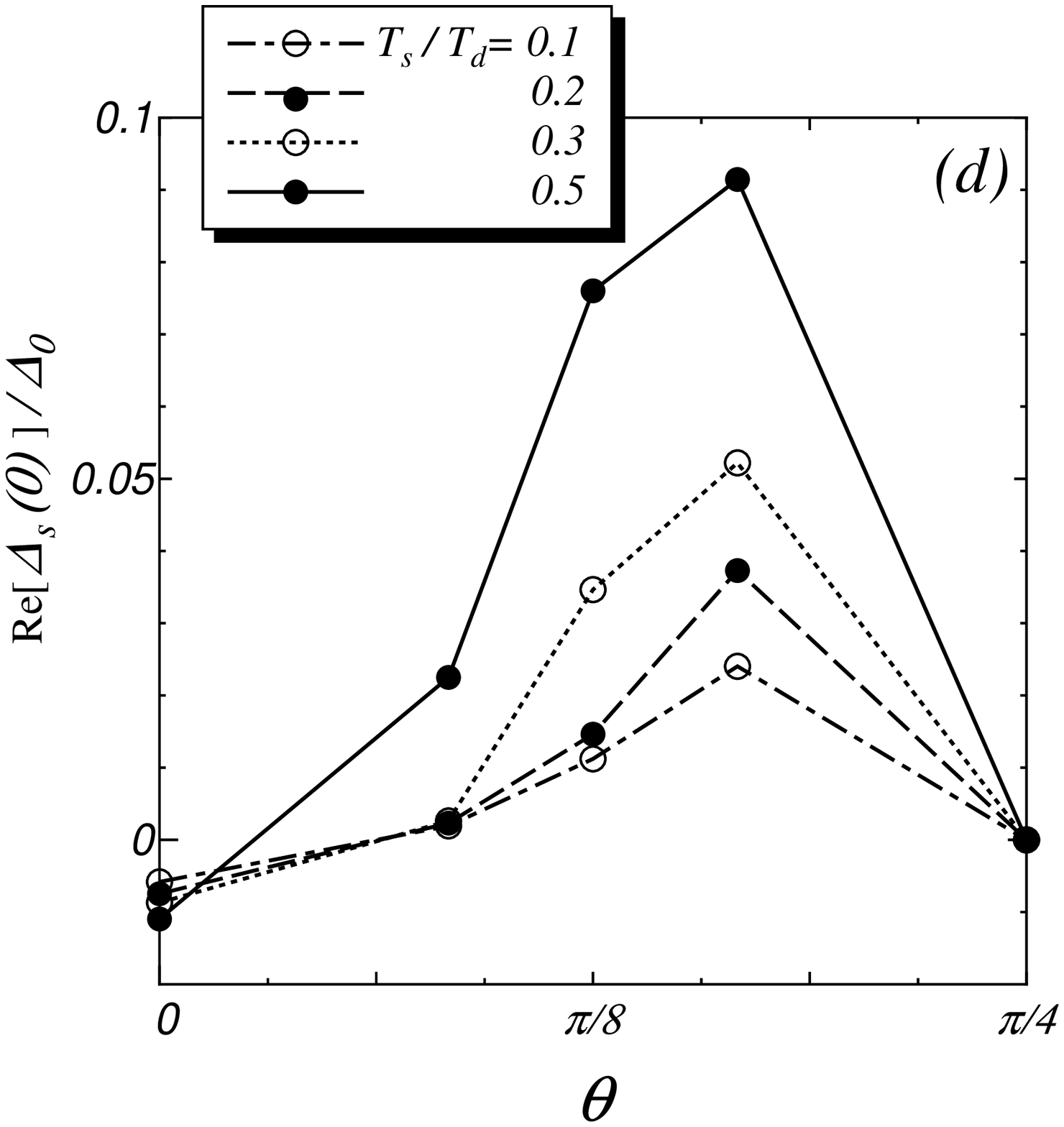}
\end{minipage}
\end{center}
\caption{
Subdominant components of the pair potentials at the 
at the interface. $T=0.05T_{c}$.
(a) $\theta =\pi/4$. 
$Z=5.0$ for (b), (c), and (d).
\label{fig03}}
\end{figure}

Next, we look at the magnitude of subdominant components 
of the pair potential at the interface,
$\mbox{Im}[\Delta_{s}(0)]$,
$\mbox{Im}[\Delta_{d}(0)]$, and
$\mbox{Re}[\Delta_{s}(0)]$,
for various $T_{s}$, $Z$, and $\theta$.
As shown in Fig.~\ref{fig03}(a),
the magnitude of $\mbox{Im}[\Delta_{s}(0)]$
increases monotonically with $T_{s}$ for
fixed $\theta$ and $Z$, and it
is enhanced for larger magnitude of $Z$.
In other words, 
the amplitude of $\mbox{Im}[\Delta_{s}(0)]$ 
is  sensitive to the transmission
probability of the junctions. 
In Fig.~\ref{fig03}(b),
$\mbox{Im}[\Delta_{s}(0)]$
is plotted as a function of $\theta$ for
sufficiently larger magnitude of $Z(=5.0)$.
For $\theta =0$, $i.e.$, junction with (100) interface,
the magnitude of
$\mbox{Im}[\Delta_{s}(0)]$ is negligibly small
near the interface 
even at the larger magnitude of $T_{s}$.
The magnitude of $\mbox{Im}[\Delta_{s}(0)]$
is a monotonically increasing function 
with the increase of  $\theta$ 
and has a maximum at $\theta=\pi/4$.
As seen from Figs.~\ref{fig03}(c) and \ref{fig03}(d), 
both the magnitude of 
$\mbox{Im}[\Delta_{d}(0)]$ and
$\mbox{Re}[\Delta_{s}(0)]$ 
is enhanced and has a maximum at a certain $\theta$.
In the intermediate $\theta$,
$i.e.$, $\theta \neq 0$ or $\theta \neq \pi/4$,
the magnitude of $\Delta(\phi_{+},x)$ and that of $\Delta(\phi_{-},x)$ 
does not coincide any more,
the interference with the quasiparticle and the pair potential
becomes complex.
Then, not only 
$\mbox{Im}[\Delta_{s}(0)]$ but also 
$\mbox{Re}[\Delta_{s}(0)]$ and $\mbox{Im}[\Delta_{d}(0)]$ 
become nonzero. 
\par
\begin{figure}[t]
\begin{center}
\begin{minipage}[t]{8cm}
\epsfxsize=8cm
\epsfbox{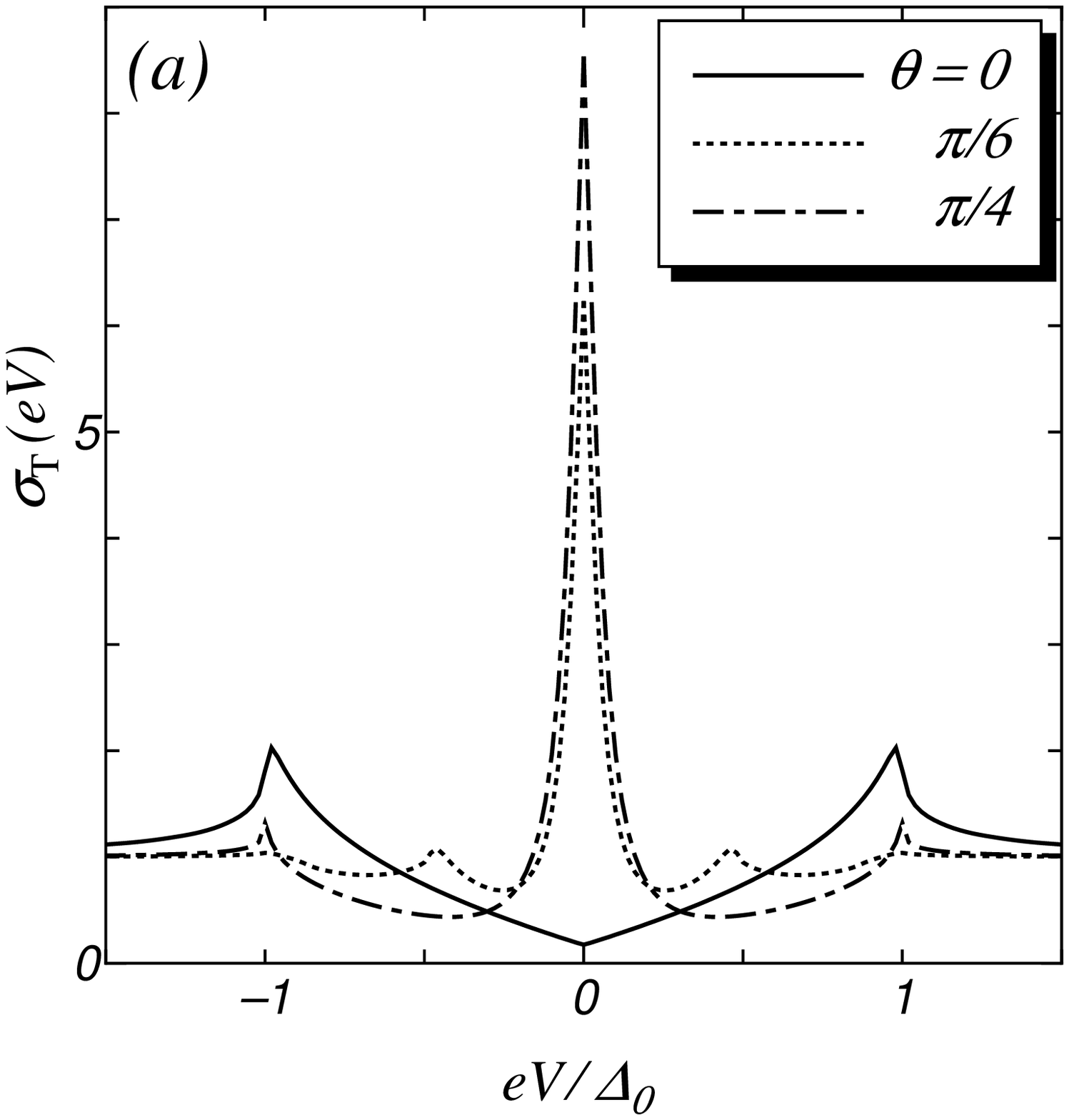}
\end{minipage}
\begin{minipage}[t]{8cm}
\epsfxsize=8cm
\epsfbox{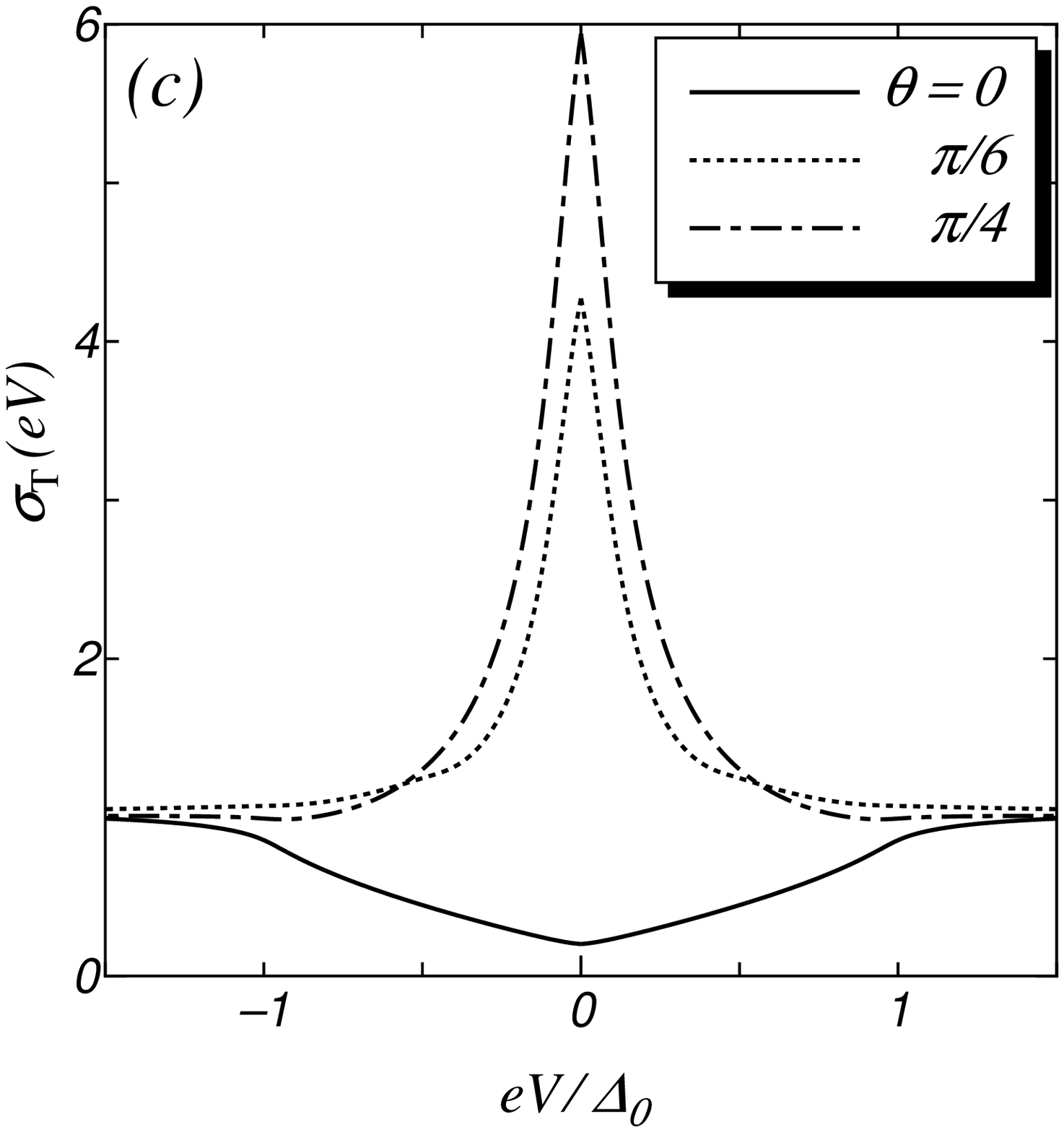}
\end{minipage}
\begin{minipage}[t]{8cm}
\epsfxsize=8cm
\epsfbox{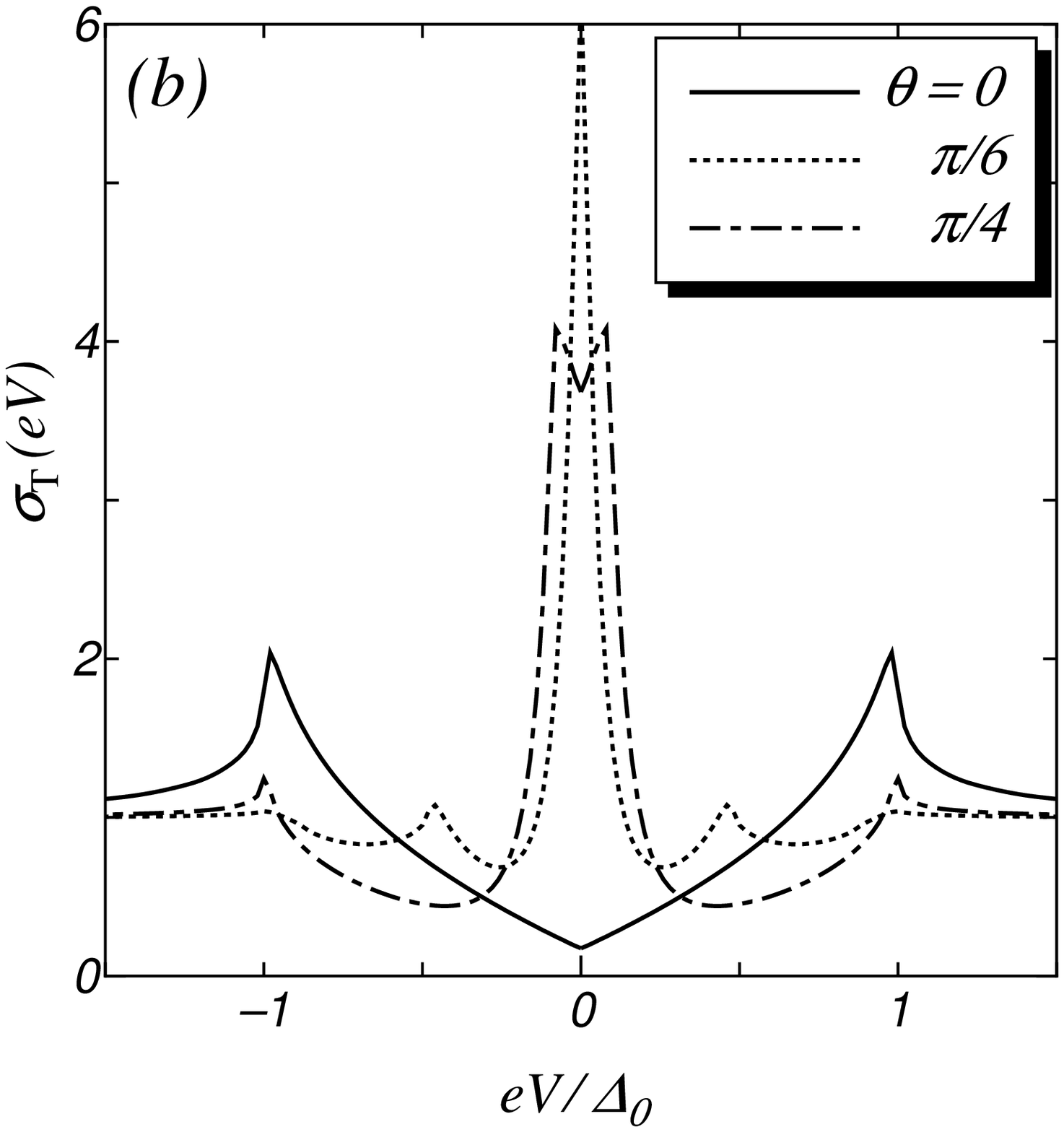}
\end{minipage}
\begin{minipage}[t]{8cm}
\epsfxsize=8cm
\epsfbox{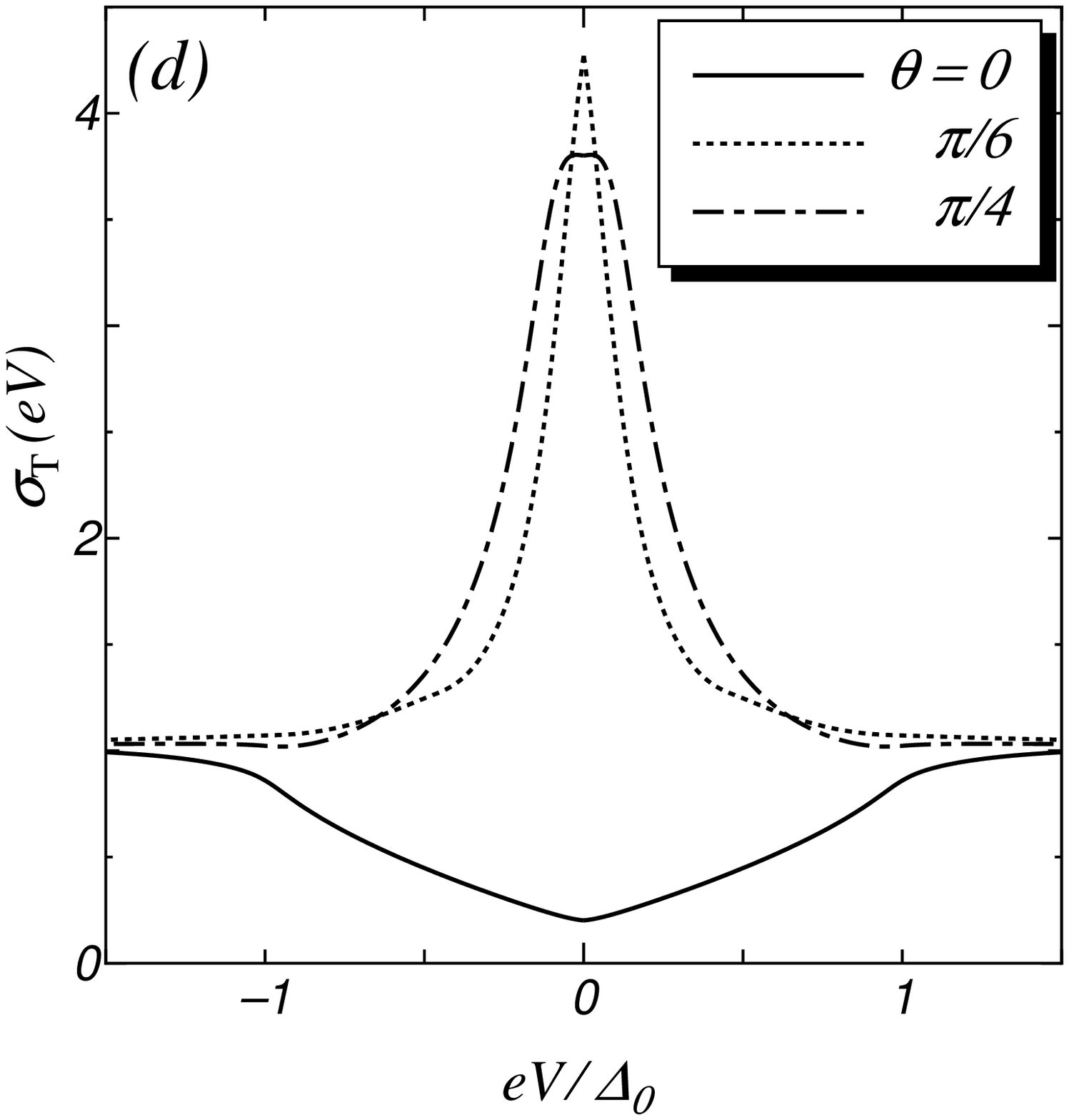}
\end{minipage}
\end{center}
\caption{
The normalized tunneling conductance
in the $d_{x^{2}-y^{2}}$+i$s$-wave state
near the interface at $Z=3.0$. 
$T_{s}/T_{d}=0.05$ for (a) and (c). 
$T_{s}/T_{d}=0.1$ for (b) and (d).
$T=0$ for (a) and (b).
$T/T_{c}=0.05$ for (c) and (d).
\label{fig04}}
\end{figure}

%
Using self-consistently determined pair potentials, 
let us look at the normalized tunneling conductance
$\sigma_{\rm T}(eV)$.
In order to clarify the temperature $T$ dependence of 
$\sigma_{\rm T}(eV)$, 
we choose $T=0$ in the left panels of Fig.~\ref{fig04} 
and $T=0.05T_{c}$ in the right panels. 
Only for $\theta=0$,
line shape of $\sigma_{\rm T}(eV)$ is similar
to that of the bulk density of states
of $d_{x^{2}-y^{2}}$-wave superconducting state. 
In other cases, 
$\sigma_{\rm T}(eV)$ has a zero bias enhanced line shape. 
As clarified in previous literatures \cite{Kasi00},
when $\theta$ deviates from zero,
since the injected and reflected quasiparticles
have a chance to feel the sign change
of the pair potentials,
zero-energy ABS is formed at the interface.
This zero-energy ABS causes the ZBCP 
when the magnitude of $T_{s}$ is small. 
With the increase of the magnitude of $T_{s}$, 
the zero energy ABS is unstable and  
$s$-wave subdominant component is induced 
which breaks time reversal symmetry and it blocks the motion of
the quasiparticles.
Then, the energy levels of bound state  shift from zero
and the local density of states 
has a zero-energy peak splitting.
The resulting $\sigma_{\rm T}(eV)$ has a 
ZBCP splitting as shown in  Fig.~\ref{fig04}(b).
However, with the increase of $T$, 
the slight splitting of ZBCP fades out
due to smearing effect by  finite temperature
and the resulting $\sigma_{\rm T}(eV)$
has a rather broad ZBCP 
[see Fig.~\ref{fig04}(c) and Fig.~\ref{fig04}(d)].
\par
\begin{figure}[htbp]
\begin{center}
\epsfxsize=15cm
\epsfysize=18.5cm
\centerline{\epsfbox{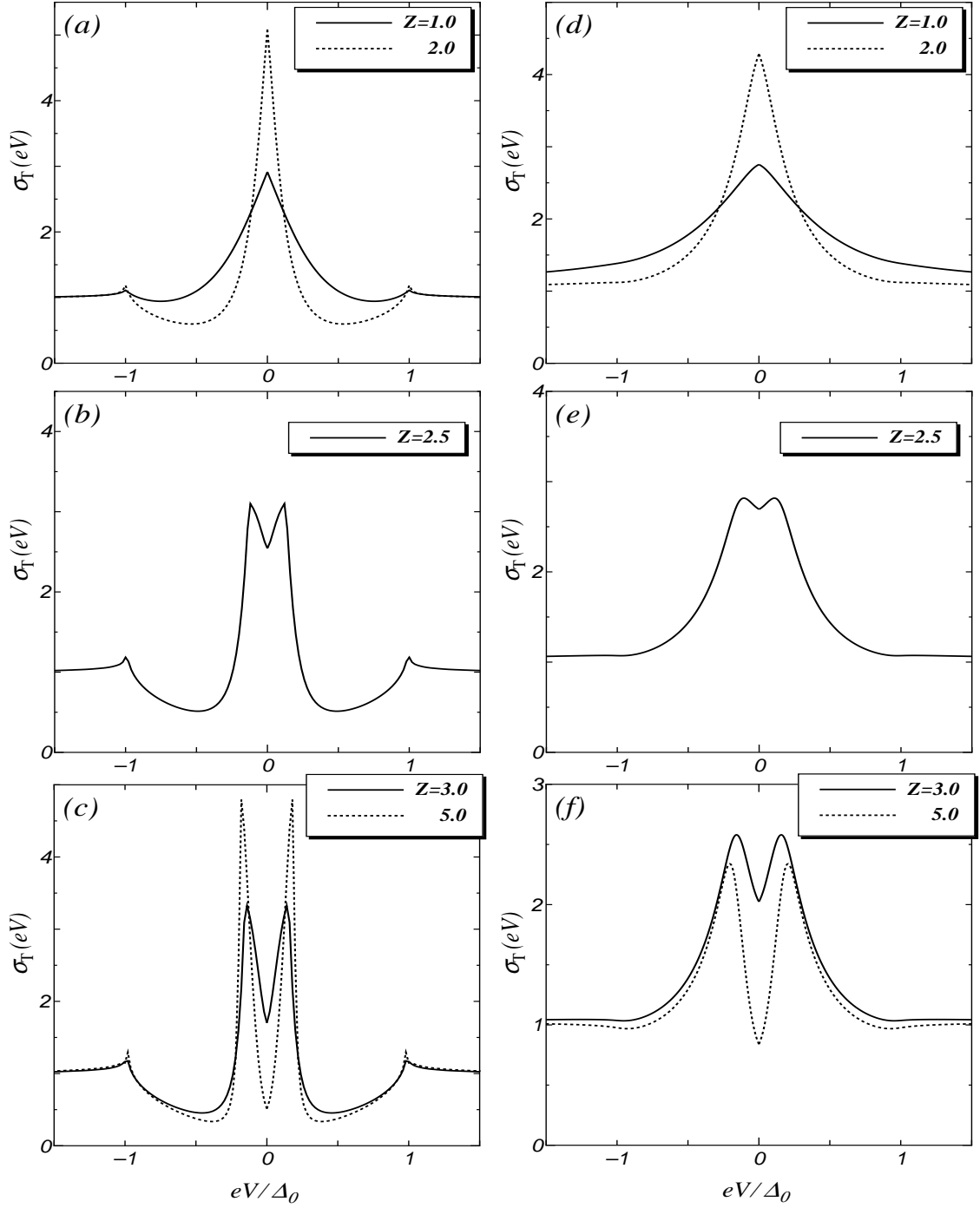}}
\caption{
Tunneling conductance
for  $d_{x^{2}-y^{2}}$+i$s$-wave state
with $T_{s}/T_{d}=0.2$ and $\theta=\pi/4$ 
for various $Z$;
(a) [(d)] low barrier ($Z=1.0$, $2.0$)
(b) [(e)] middle barrier ($Z=2.5$),
and
(c) [(f)] high barier ($Z=3.0$, $5.0$).
$T=0$ for (a), (b), and (c). 
$T/T_{c}=0.05$ for (d), (e), and (f).
\label{fig05}}
\end{center}
\end{figure}

%
Next, we concentrate on 
how $\sigma_{\rm T}(eV)$ is influenced
by the transmission probability of the junctions,
$i.e.$, the magnitude of $Z$.
In Fig.~\ref{fig05}, $\sigma_{\rm T}(eV)$ with $\theta=\pi/4$ is
plotted for $T=0$ (left panels) and $T=0.05T_{c}$ (right panels).
For the junctions with high transmissivity,
$\sigma_{\rm T}(eV)$ has a ZBCP, [see Fig.~\ref{fig05}(a)]
and the magnitude of $\sigma_{\rm T}(0)$
is firstly enhanced with the increase of $Z$.
In this case, the predominant
$d_{x^{2}-y^{2}}$-wave component only
exists near the interface
as shown in Fig.~\ref{fig01}(b). 
However, with the increase of $Z$,
$\sigma_{\rm T}(eV)$ starts to have a
ZBCP splitting at a certain value of $Z$, 
where the magnitude of subdominant component $\mbox{Im}[\Delta_{s}(x)]$
at the interface becomes the same order as that of the predominant 
component $\mbox{Re}[\Delta_{d}(x)]$.
For sufficiently larger magnitude of $Z$, 
$\sigma_{\rm T}(eV)$ has a  ZBCP splitting,
[see Fig.~\ref{fig05}(b) and \ref{fig05}(c),] 
and the magnitude of $\sigma_{\rm T}(0)$ 
decreases with the increase of $Z$.
However, the above obtained results are 
influenced by finite temperature effect. 
The right panels of Fig.~\ref{fig05} is shown
for the tunneling conductance in $T/T_{c}=0.05$.
The slight enhanced structure of $\sigma_{T}(eV)$ 
at $eV= \pm \Delta_{0}$ in 
Fig.~\ref{fig05}(a),
\ref{fig05}(b), and \ref{fig05}(c) 
is invisible due to the smearing effect
by finite temperature  [see Fig.~\ref{fig05}(c),
\ref{fig05}(d), and \ref{fig05}(e)]. 
With the increase of the magnitude of $Z$, 
$\sigma_{\rm T}(eV)$ has a  ZBCP with tiny dip 
even at $Z=2.5$,
where the order of the amplitude of $\mbox{Im}[\Delta_{s}(0)]$
is 0.2$\Delta_{0}$.
With the further increase of $Z$,
$\sigma_{\rm T}(eV)$ has a ZBCP splitting [see Fig.~\ref{fig05}(f)],
however the degree of the splitting is significantly weakened
as compared to the corresponding curves in   Fig.~\ref{fig05}(c). 
\par
\begin{figure}[htbp]
\begin{center}
\epsfxsize=7cm
\epsfysize=18cm
\centerline{\epsfbox{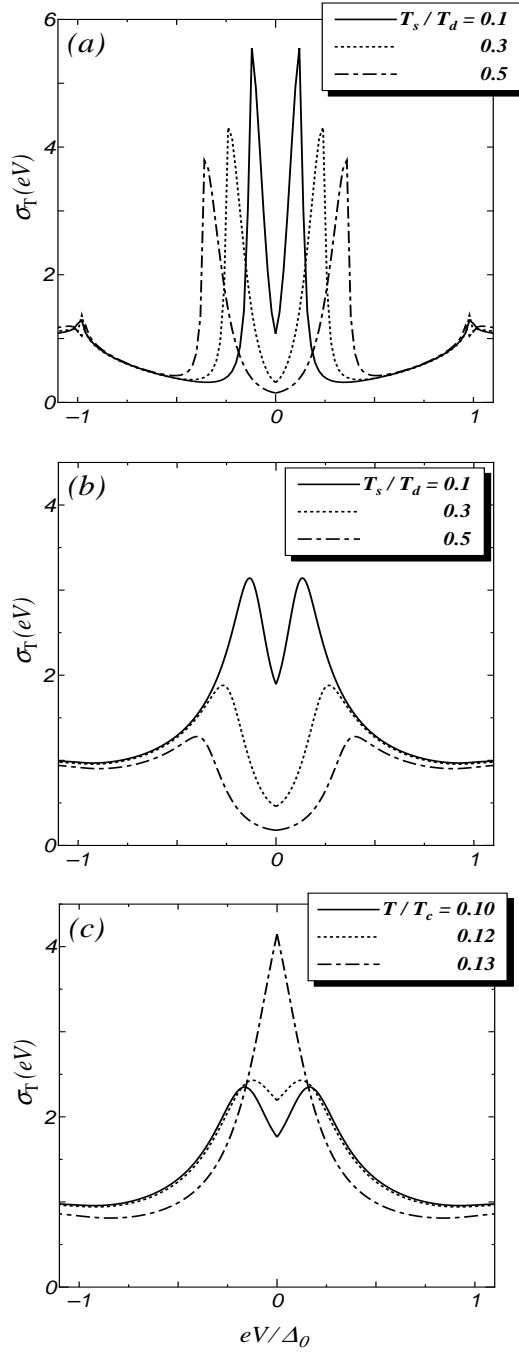}}
\caption{
Tunneling conductance 
for $d_{x^{2}-y^{2}}$+i$s$-wave state. 
$Z=5$ and $\theta=\pi/4$. 
(a) $T=0$,  
(b) $T=0.05T_{c}$ 
for various  magnitude of $T_{s}$. 
(c)  various temperature with $T_{s}/T_{d}=0.2$.
\label{fig06}}
\end{center}
\end{figure}

Finally, we  look at the relation
between the position of the splitted peak and
the magnitude of $\mbox{Im}[\Delta_{s}(0)]$.
In Fig.~\ref{fig06}(a),
$\sigma_{\rm T}(eV)$ is plotted for various 
$T_{s}$ with $Z=5$ and $T=0$.
As shown in Fig.~\ref{fig03}(b), 
the magnitude of the induced subdominant imaginary
component of $\mbox{Im}[\Delta_{s}(0)]$ is about
$0.16\Delta_{0}$, $0.3\Delta_{0}$, and 
$0.42\Delta_{0}$ 
for $T_{s}/T_{d}=0.1$, $T_{s}/T_{d}=0.3$
and $T_{s}/T_{d}=0.5$, respectively.
The corresponding $\sigma_{\rm T}(eV)$ has a splitted peak
locating at $\pm 0.16\Delta_{0}$, $\pm 0.3\Delta_{0}$,
and $\pm 0.42\Delta_{0}$, respectively.
With the increase of $T$, the height of these
peaks are drastically suppressed
as shown in Fig.~\ref{fig06}(b).
Finally, we show how the line shape of
$\sigma_{\rm T}(eV)$ changes at the temperature $T=T_{\bar{s}}$ 
($T_{\bar{s}}=0.12T_{c}$) 
where $\mbox{Im}[\Delta_{s}(x)]$ becomes nonzero.
As seen from Fig.~\ref{fig06}(c), the magnitude of 
$\sigma_{T}(0)$ is reduced with the introduction of 
$\mbox{Im}[\Delta_{s}(x)]$. \par
At the end of this subsection, 
we can summarize that 
even in the presence of BTRSS, 
the resulting $\sigma_{\rm T}(eV)$
does not always have a clear ZBCP splitting 
due to the finite temperature effect 
when  the  magnitude of the transmission probability
of the junctions is not low. 
\par
\subsection{$d_{x^{2}-y^{2}}${\rm +i}$d_{xy}$-wave state}
\begin{figure}[t]
\begin{center}
\begin{minipage}[t]{8cm}
\epsfxsize=8cm
\epsfbox{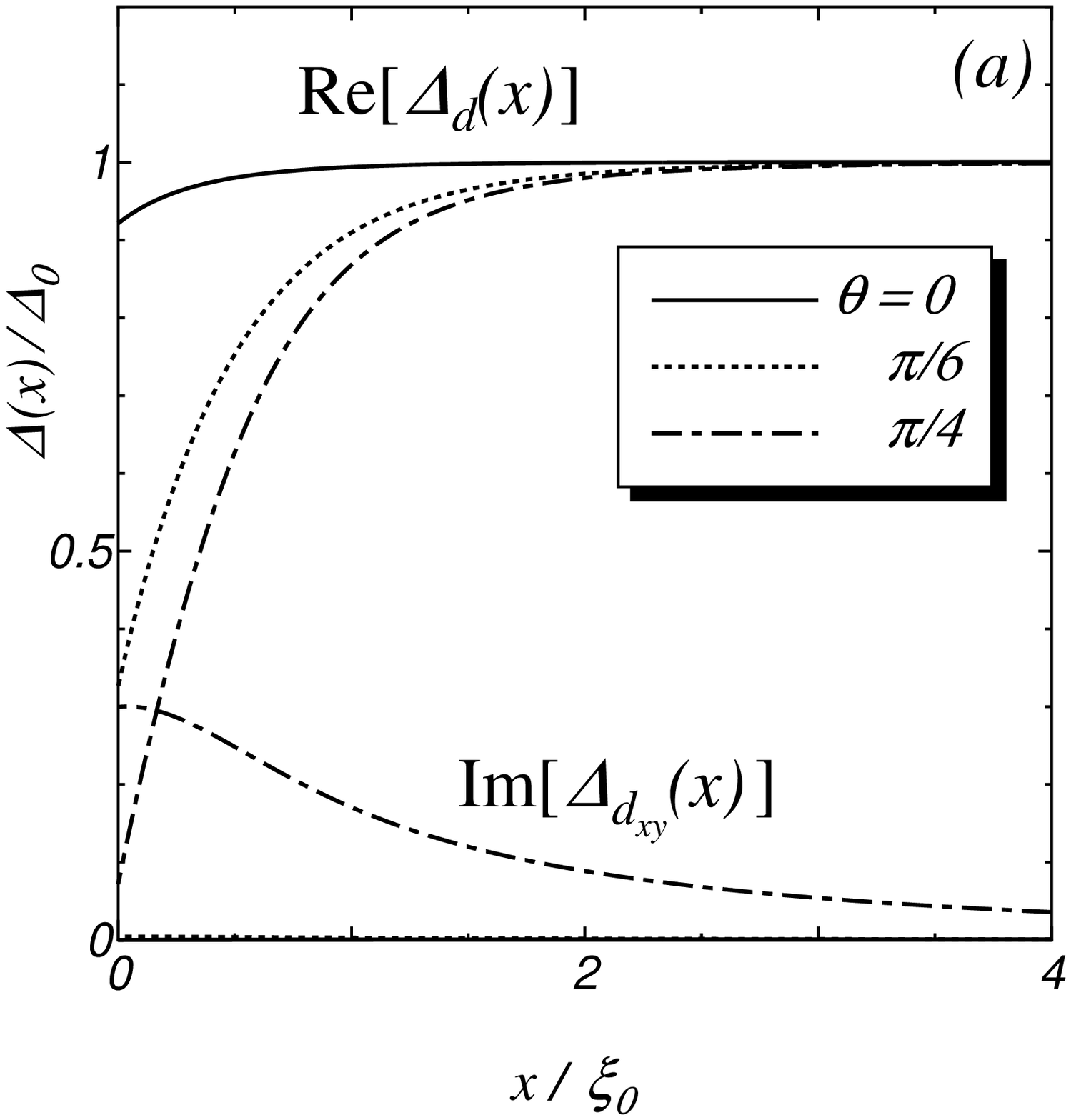}
\end{minipage}
\begin{minipage}[t]{8cm}
\epsfxsize=8cm
\epsfbox{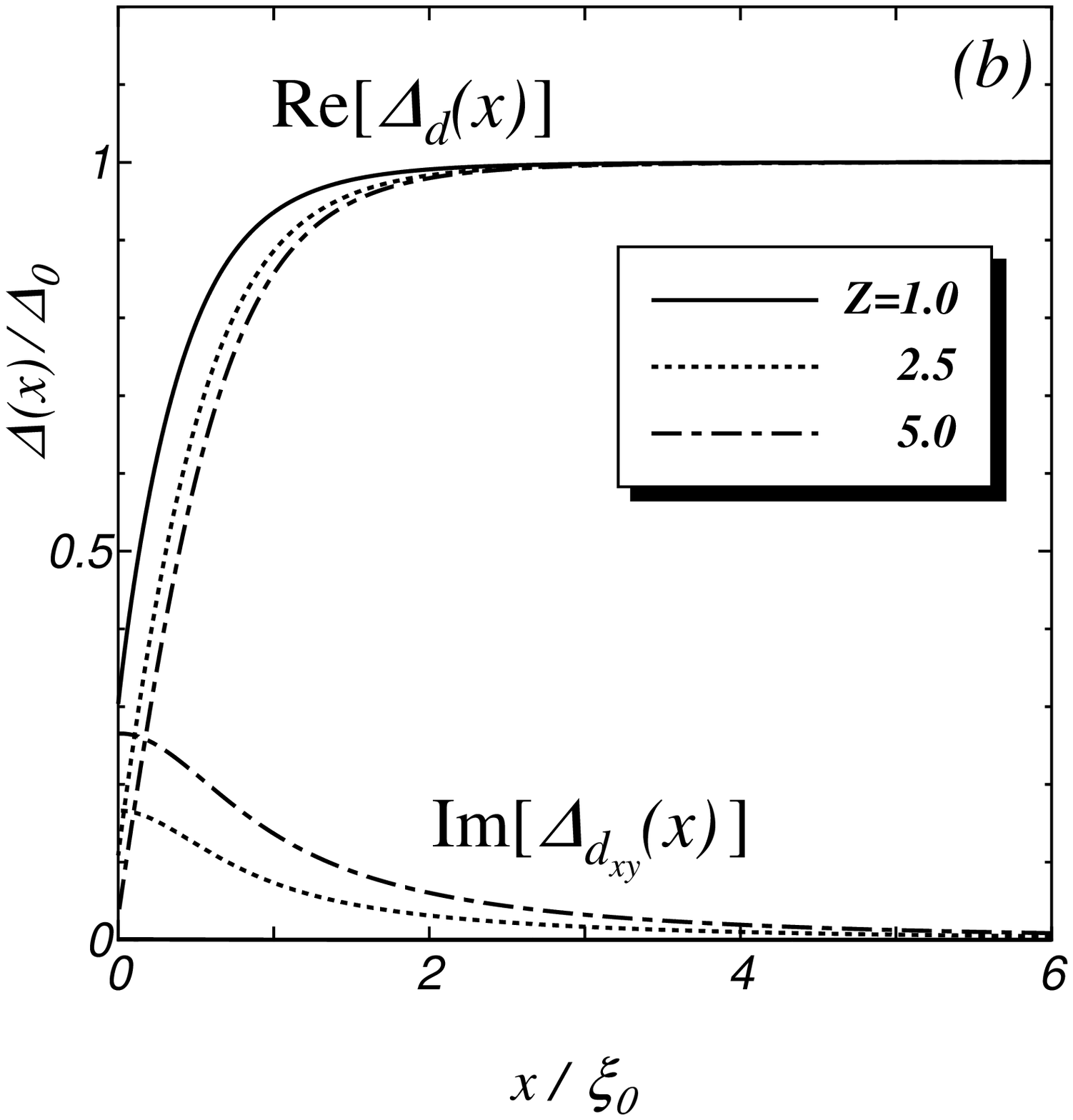}
\end{minipage}
\end{center}
\caption{Spatial dependences of the pair potential 
with $T_{d_{xy}}/T_{d}=0.2$. $T=0.05T_{c}$. 
(a) $Z=3.0$ for various $\theta$. 
(b) $\theta=0$ for various $Z$. \label{fig07}}
\end{figure}

In this subsection,
we study spatial dependence of the pair potentials of
the $d_{x^{2}-y^{2}}$+i$d_{xy}$-wave state
and the resulting tunneling conductance
in  N/D junctions.
The pair potential is given by
\begin{eqnarray}
\Delta(\phi,x)=\Delta_{d}(x)\cos [2(\phi-\theta)]
+\Delta_{d_{xy}}(x)\sin [2(\phi-\theta)],
\end{eqnarray}
where $\Delta_{d_{xy}}(x)$
is an amplitude of
the $d_{xy}$-wave superconducting state
and a complex number. 
The attractive inter-electron potential
$V(\phi,\phi^{\prime})$ is given by 
\begin{eqnarray}
V(\phi,\phi^{\prime})=
2V_{d}\cos[2(\phi-\theta)]\cos[2(\phi^{\prime}-\theta)] 
+ 2V_{d_{xy}}\sin[2(\phi-\theta)]\sin[2(\phi^{\prime}-\theta)],
\end{eqnarray}
where $V_{d}$ and $V_{d_{xy}}$ stand for the attractive
potential of predominant $d_{x^{2}-y^{2}}$-wave
and subdominant $d_{xy}$-wave, respectively,
and they are given as
\begin{eqnarray}
V_{d} &=& \frac{2\pi k_{\rm B}T}
{\displaystyle \log \frac{T}{T_{c}}
+ \sum_{0 \leq m < \omega_{c}/2\pi T}\frac{1}{m+1/2} }, 
\\
V_{d_{xy}} &=& \frac{2\pi k_{\rm B}T}
{\displaystyle \log \frac{T}{T_{d_{xy}}}
+ \sum_{0 \leq m < \omega_{c}/2\pi T}\frac{1}{m+1/2} }. 
\end{eqnarray}
The spatial dependence of the pair potentials
with a finite transmissivity for $Z=3.0$ and $\theta=\pi/4$
is shown in Fig.~\ref{fig07}(a).
As in the case for  $d_{x^{2}-y^{2}}$+i$s$-wave state,
the  amplitude of $\mbox{Im}[\Delta_{d_{xy}}(x)]$ vanishes 
at $\theta =0$. 
For $\theta=\pi/4$,
the suppression of the magnitude of
$\mbox{Re}[\Delta_{d}(x)]$ is most significant,
while $\Delta_{d_{xy}}(x)$ is
induced at the interface.
At this $\theta$, 
the $d_{xy}$-wave component
is not affected by depairing effect seriously
since the lobe direction of
$d_{xy}$-wave pair potential is parallel or perpendicular
to the interface as in the case for predominant 
$d_{x^{2}-y^{2}}$-wave with $\theta=0$.
As shown in  Fig.~\ref{fig07}(b),
the amplitude of $\mbox{Im}[\Delta_{d_{xy}}(x)]$
is enhanced with the increase of $Z$.
\par
\begin{figure}[t]
\begin{center}
\begin{minipage}[t]{8cm}
\epsfxsize=8cm
\epsfbox{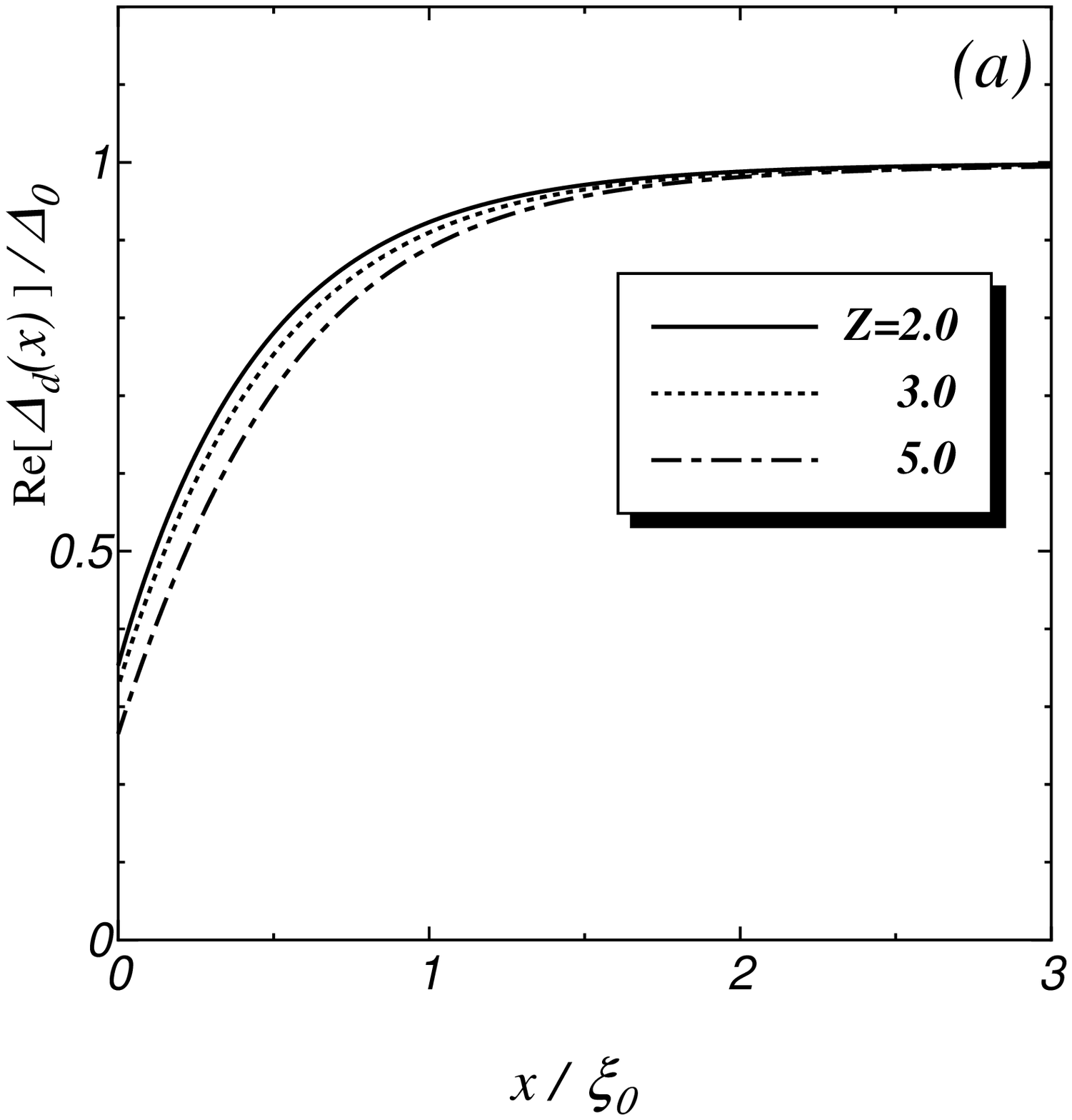}
\end{minipage}
\begin{minipage}[t]{8cm}
\epsfxsize=8cm
\epsfbox{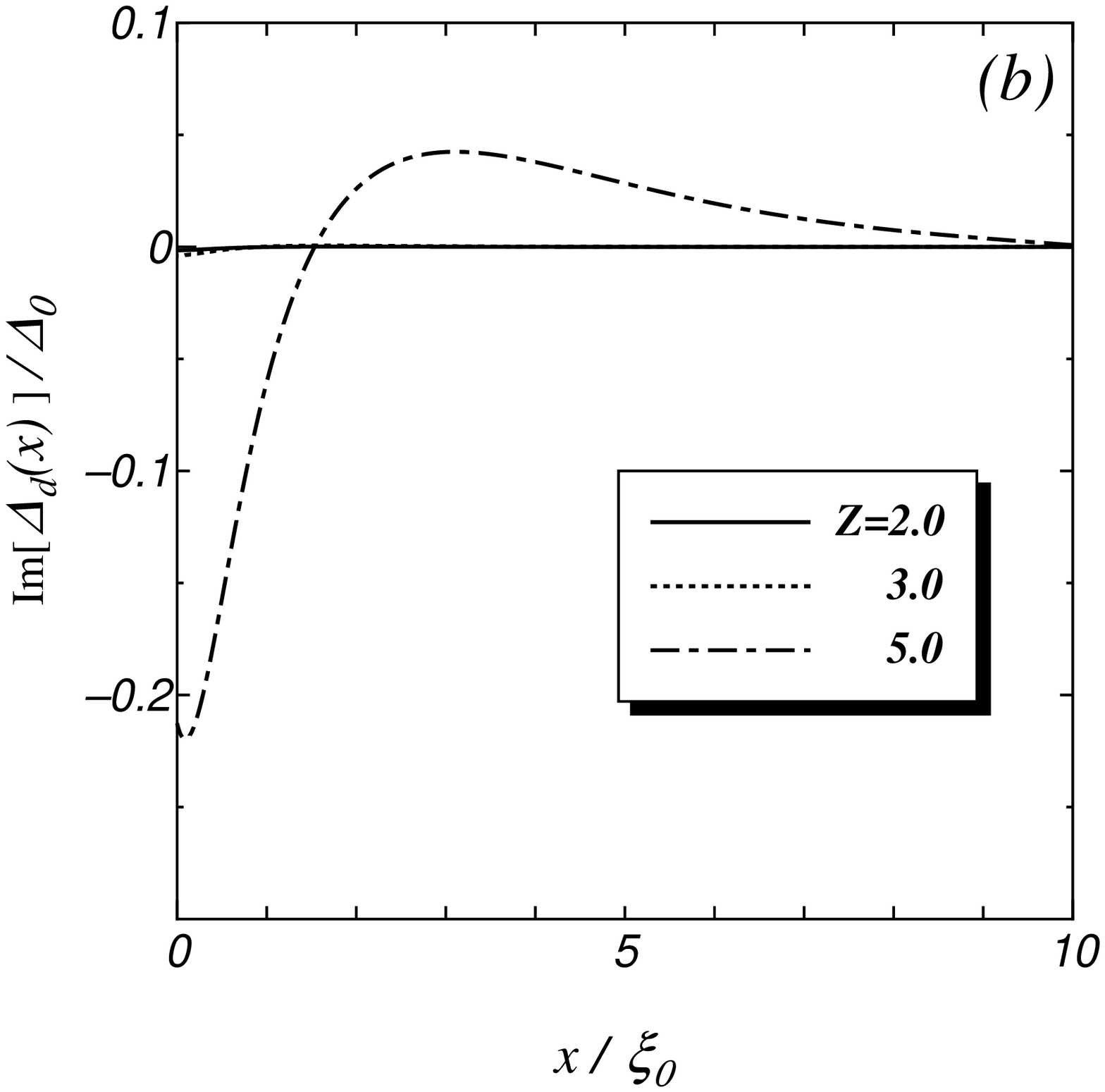}
\end{minipage}
\begin{minipage}[t]{8cm}
\epsfxsize=8cm
\epsfbox{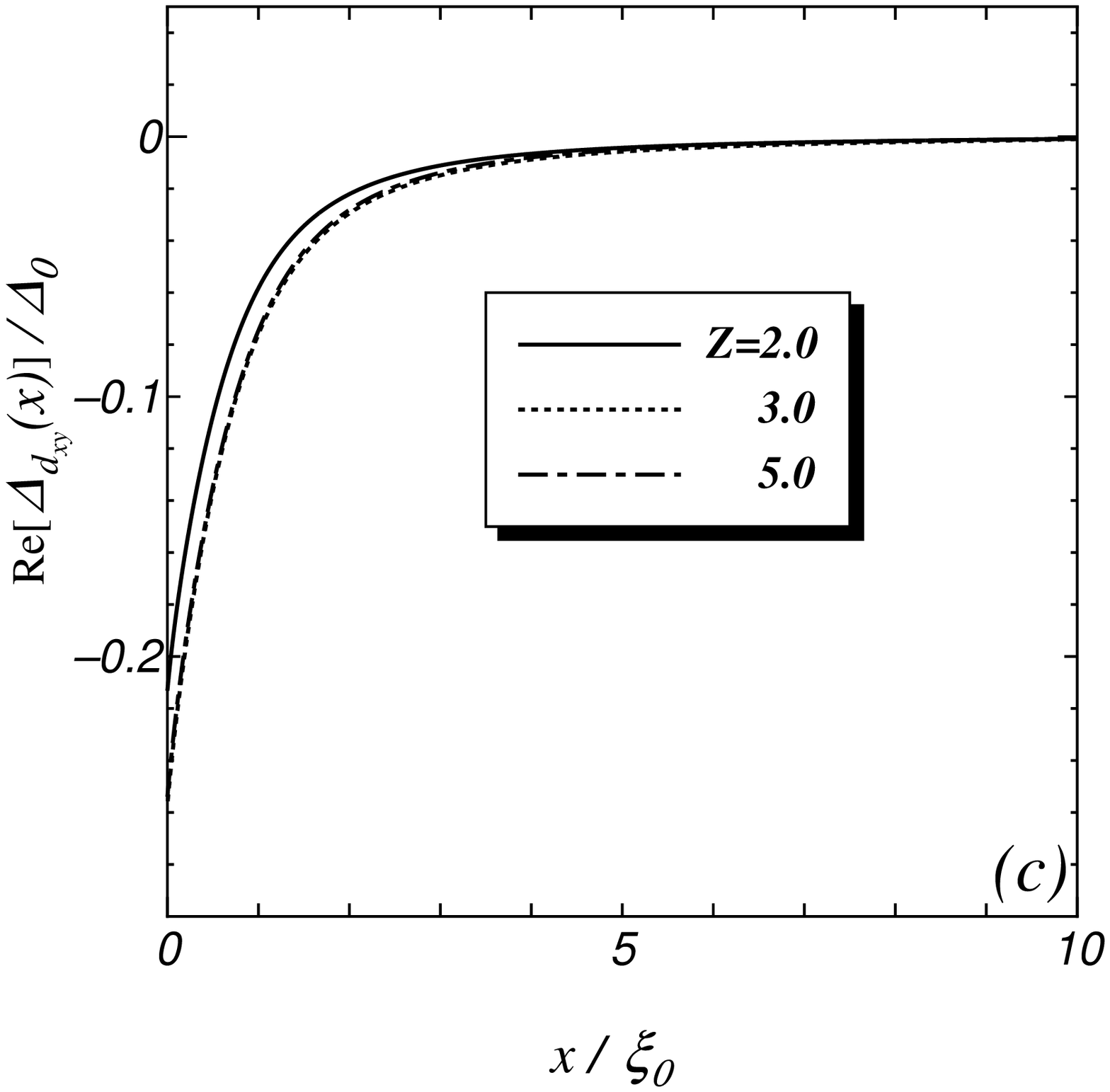}
\end{minipage}
\begin{minipage}[t]{8cm}
\epsfxsize=8cm
\epsfbox{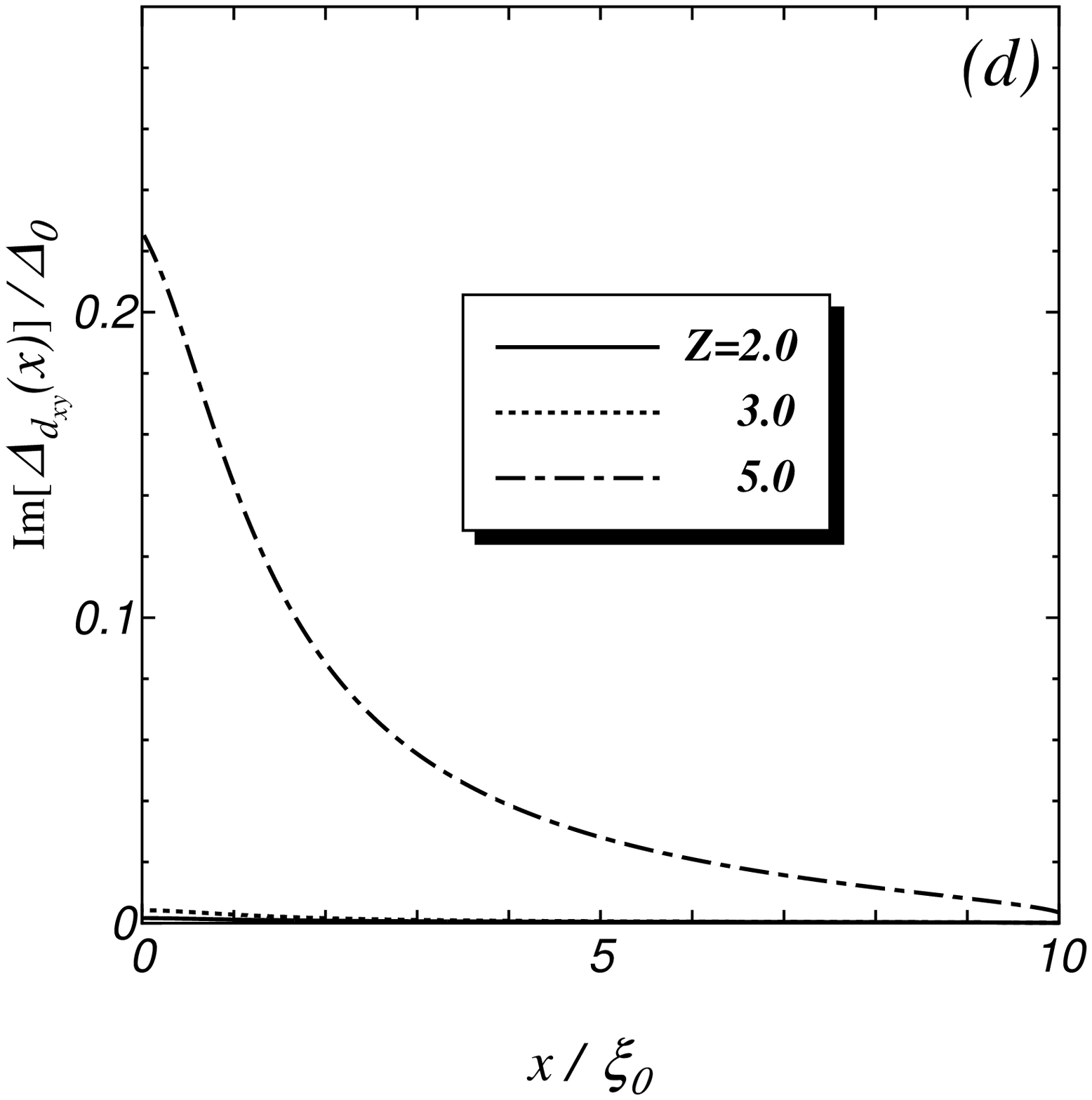}
\end{minipage}
\end{center}
\caption{
Spatial dependence of the pair potentials
near the interface with $\theta =\pi/6$
and $T_{d_{xy}}/T_{d}=0.3$;
(a) real part of $\Delta_{d}(x)$,
(b) imaginary part of $\Delta_{d}(x)$,
(c) real part of $\Delta_{d_{xy}}(x)$,
and
(d) imaginary part of $\Delta_{d_{xy}}(x)$.
$T=0.02T_{c}$.
\label{fig08}}
\end{figure}

%
The spatial dependencies of
$\mbox{Re}[\Delta_{d}(x)]$,
$\mbox{Im}[\Delta_{d}(x)]$,
$\mbox{Re}[\Delta_{s}(x)]$,
and $\mbox{Im}[\Delta_{s}(x)]$
are plotted in Figs.~\ref{fig08}(a),
\ref{fig08}(b), \ref{fig08}(c), and \ref{fig08}(d),
respectively, with the intermediate $\theta$
($\theta =\pi/6$) for various $Z$.
In such a case, the spatial dependence is much more complex as 
compared to that for $\theta=0$ or $\theta=\pi/4$, and 
both $\mbox{Im}[\Delta_{d}(x)]$ and
$\mbox{Re}[\Delta_{d_{xy}}(x)]$ become nonzero. 
The amplitudes of
$\mbox{Im}[\Delta_{d}(x)]$,
$\mbox{Re}[\Delta_{d_{xy}}(x)]$,
and $\mbox{Im}[\Delta_{s}(x)]$
are enhanced for larger magnitude of $Z$,
where the suppression of
$\mbox{Re}[\Delta_{d}(x)]$ is significant.
The remarkable feature is
that the amplitude of 
$\mbox{Im}[\Delta_{d}(x)]$ and
$\mbox{Re}[\Delta_{d_{xy}}(x)]$
can become the same order of
$\mbox{Im}[\Delta_{s}(x)]$ for larger $Z$.
\par
\begin{figure}[t]
\begin{center}
\begin{minipage}[t]{8cm}
\epsfxsize=8cm
\epsfbox{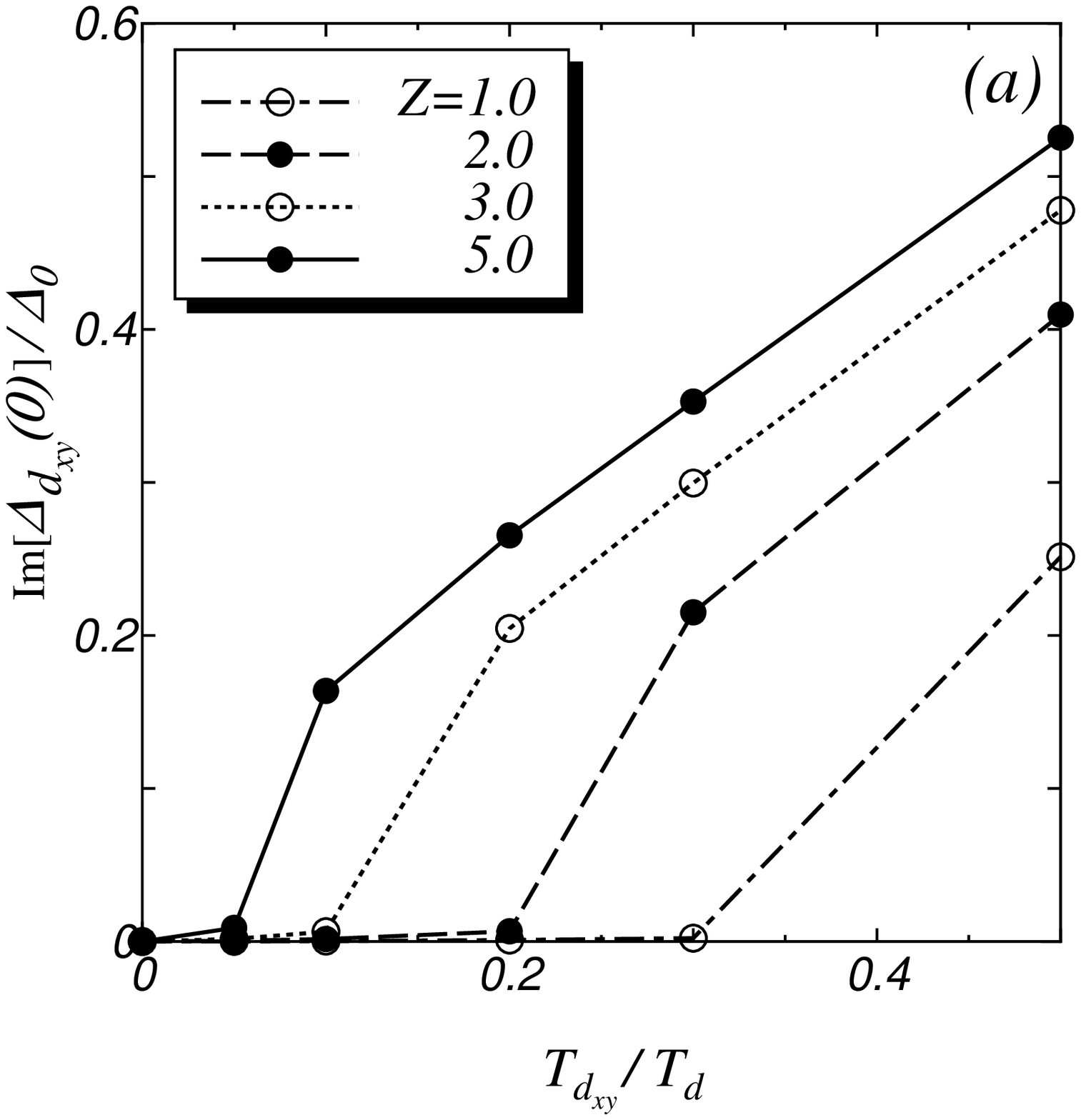}
\end{minipage}
\begin{minipage}[t]{8cm}
\epsfxsize=8cm
\epsfbox{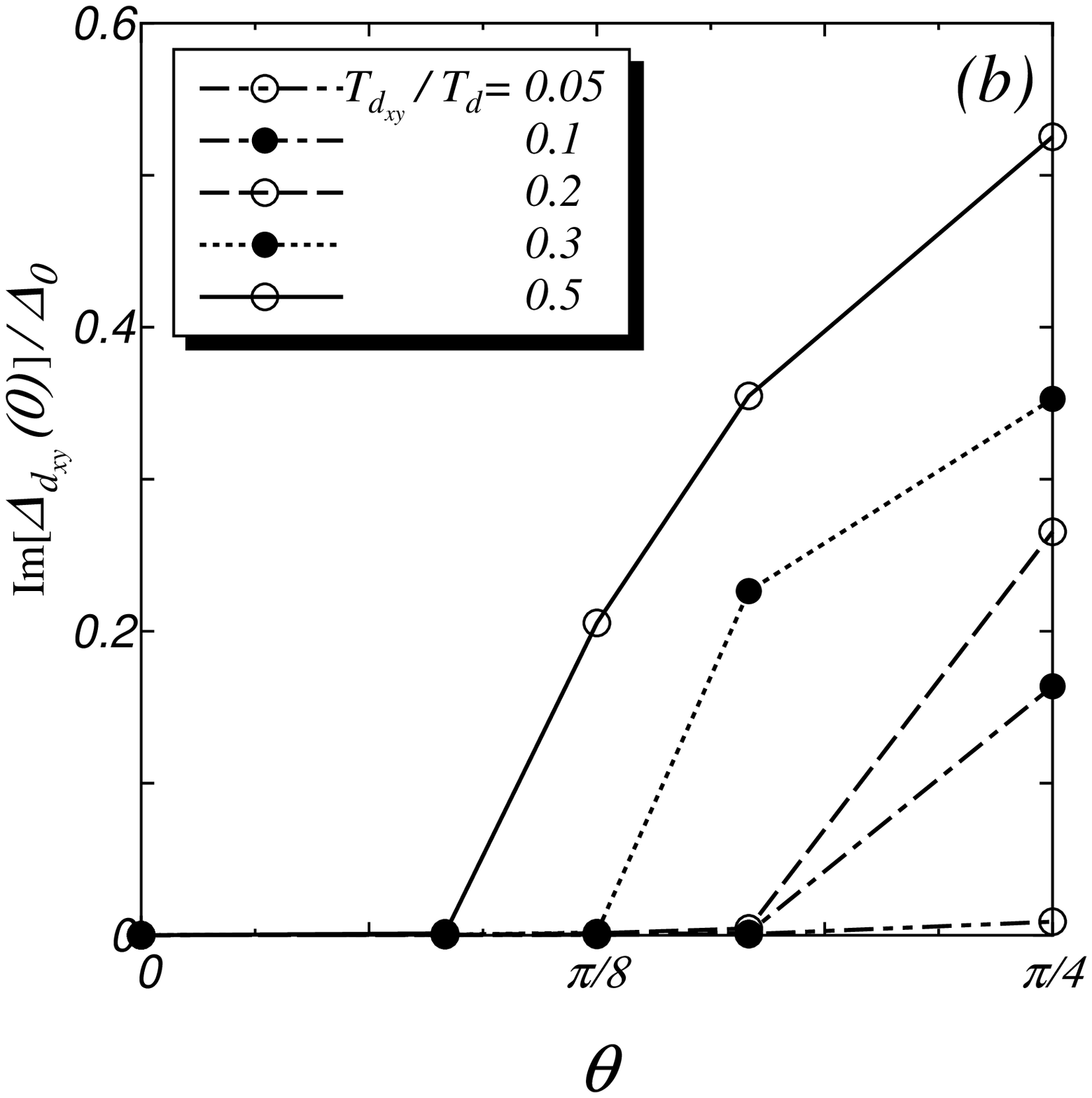}
\end{minipage}
\begin{minipage}[t]{8cm}
\epsfxsize=8cm
\epsfbox{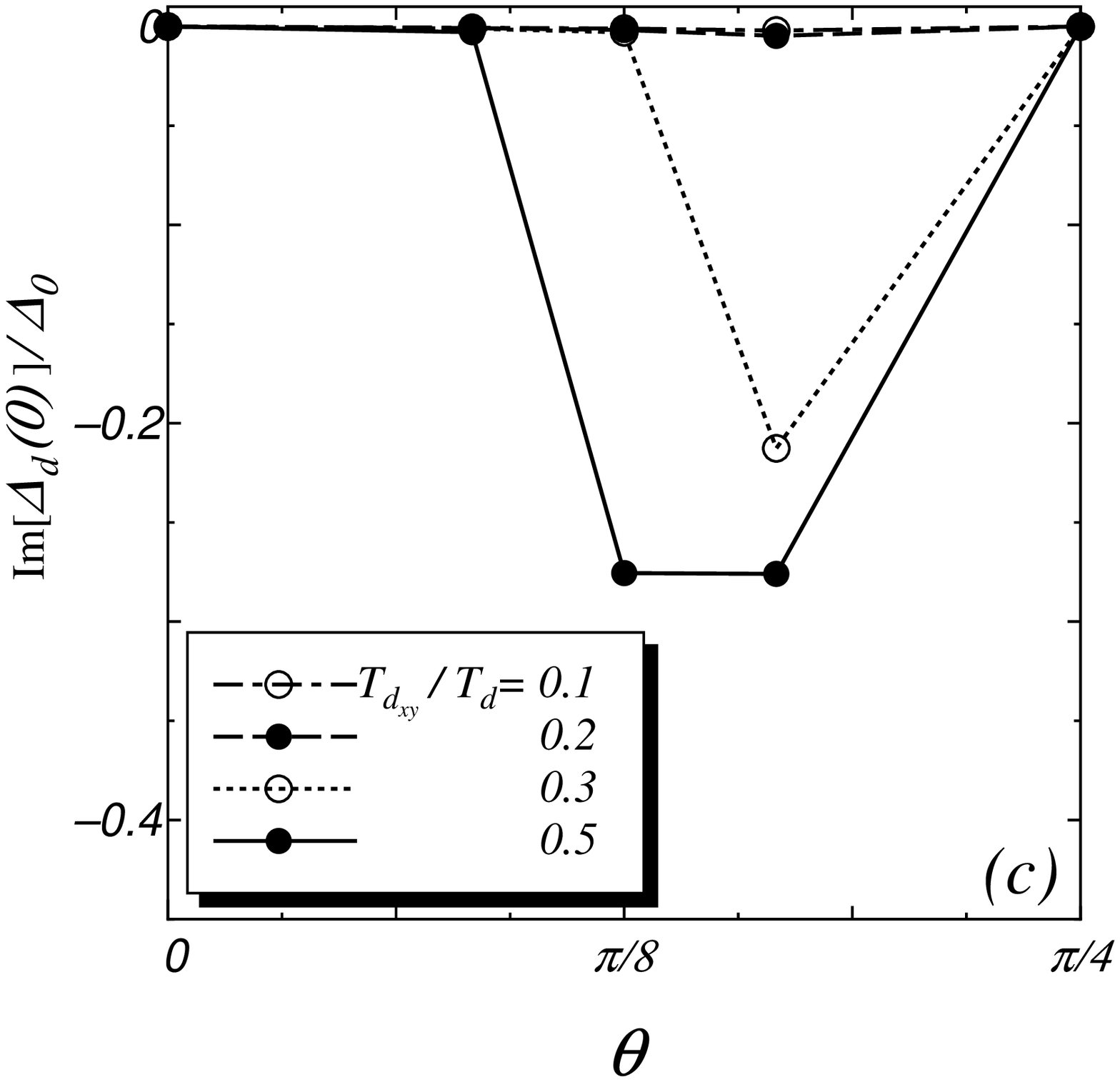}
\end{minipage}
\begin{minipage}[t]{8cm}
\epsfxsize=8cm
\epsfbox{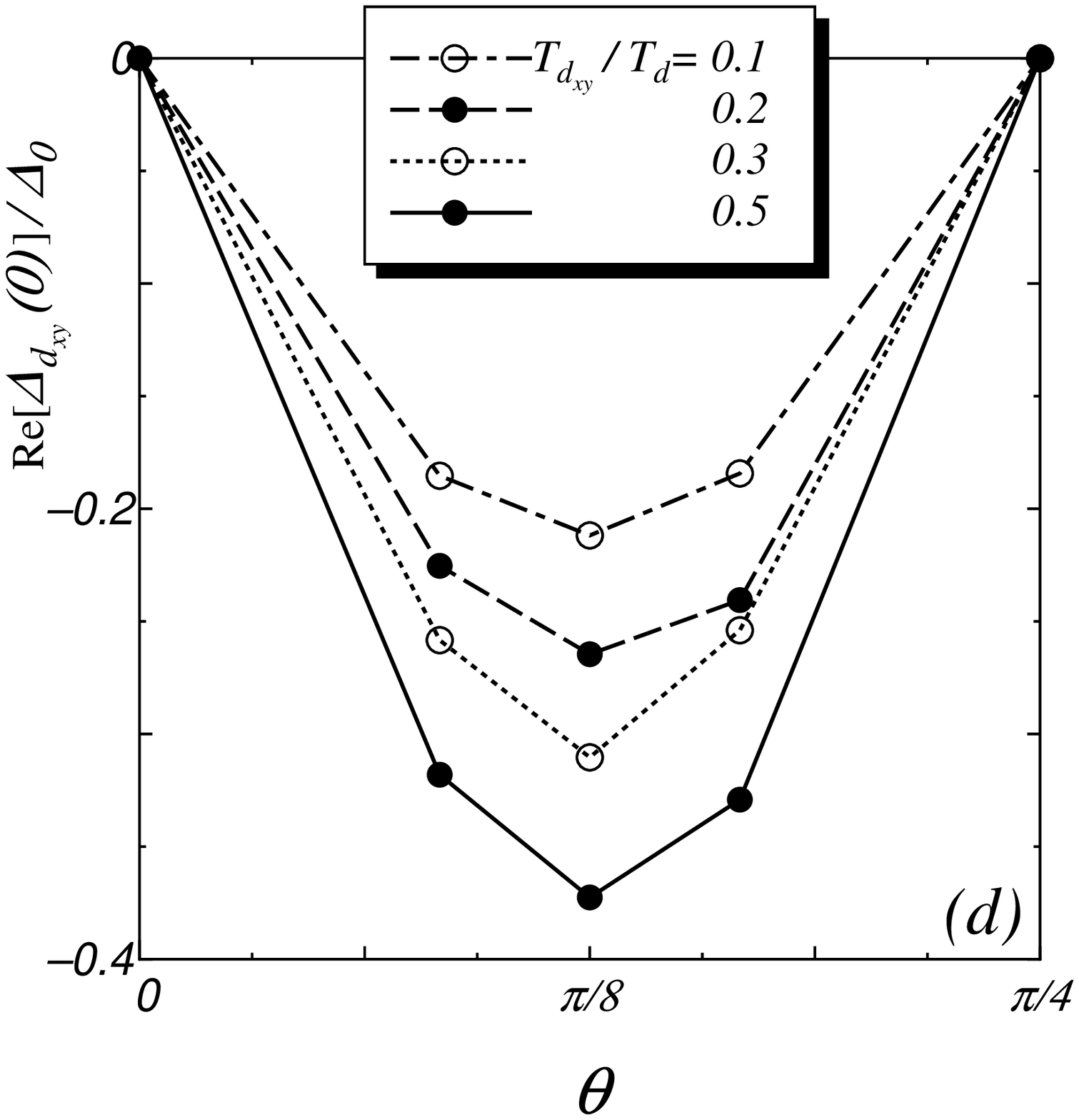}
\end{minipage}
\end{center}
\caption{
Subdominant components of the pair potentials at the 
interface. $T=0.05T_{c}$. 
(a) $\theta =\pi/4$. 
$Z=5.0$ for (b), (c), and (d). 
\label{fig09}}
\end{figure}

Next, we look at the magnitude of subdominant components
of the pair potential at the interface,
$\mbox{Im}[\Delta_{d_{xy}}(0)]$,
$\mbox{Im}[\Delta_{d}(0)]$,
and
$\mbox{Re}[\Delta_{d_{xy}}(0)]$
for various parameters $T_{d_{xy}}$, $Z$
and $\theta$
shown in Fig.~\ref{fig09}.
The magnitude of $\mbox{Im}[\Delta_{d_{xy}}(0)]$ 
increases monotonically with $T_{d_{xy}}$
for fixed $\theta$ and  $Z$.
It is also enhanced for larger magnitude of $Z$.
Comparing the corresponding situation of
$d_{x^{2}-y^{2}}+{\rm i}s$-wave shown in Fig.~\ref{fig03}(a), 
the magnitude of $\mbox{Im}[\Delta_{d_{xy}}(0)]$ is suppressed 
for the small magnitude of $T_{d_{xy}}$. 
In Fig.~\ref{fig09}(b),
$\mbox{Im}[\Delta_{d_{xy}}(0)]$
is plotted as a function of $\theta$
for sufficiently larger magnitude of $Z(=5.0)$.
The magnitude of $\mbox{Im}[\Delta_{d_{xy}}(0)]$
is a monotonically increasing function
with the increase of $\theta$
and has a maximum at $\theta=\pi/4$.
The amplitudes of $\mbox{Im}[\Delta_{d}(0)]$ and
$\mbox{Re}[\Delta_{d_{xy}}(0)]$ are negligibly small
for $\theta=0$ and $\theta=\pi/4$,
and has a maximum at a certain $\theta$ as 
shown in Figs.~\ref{fig09}(c) and \ref{fig09}(d),
respectively.
\par
\begin{figure}[t]
\begin{center}
\begin{minipage}[t]{8cm}
\epsfxsize=8cm
\epsfbox{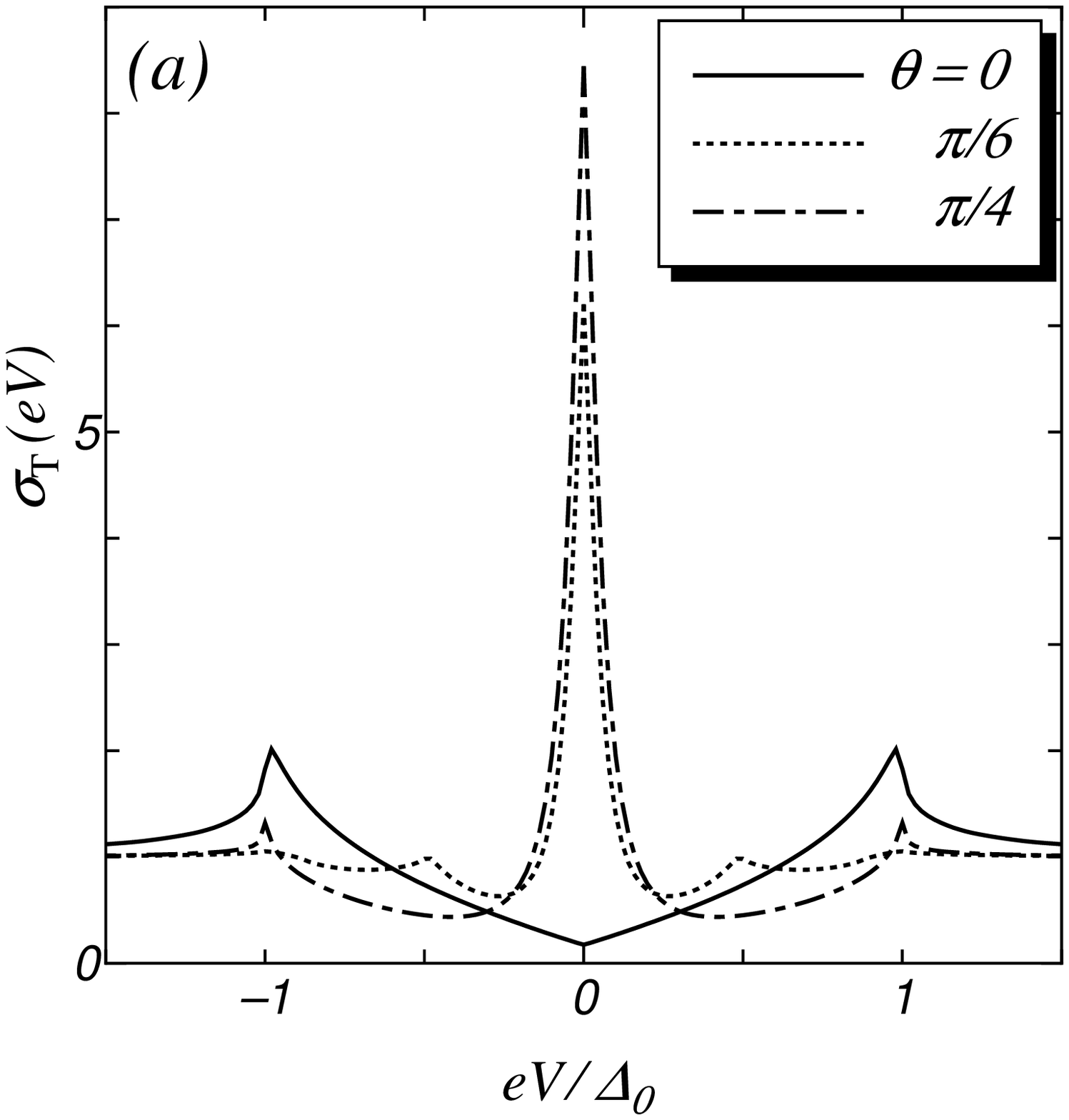}
\end{minipage}
\begin{minipage}[t]{8cm}
\epsfxsize=8cm
\epsfbox{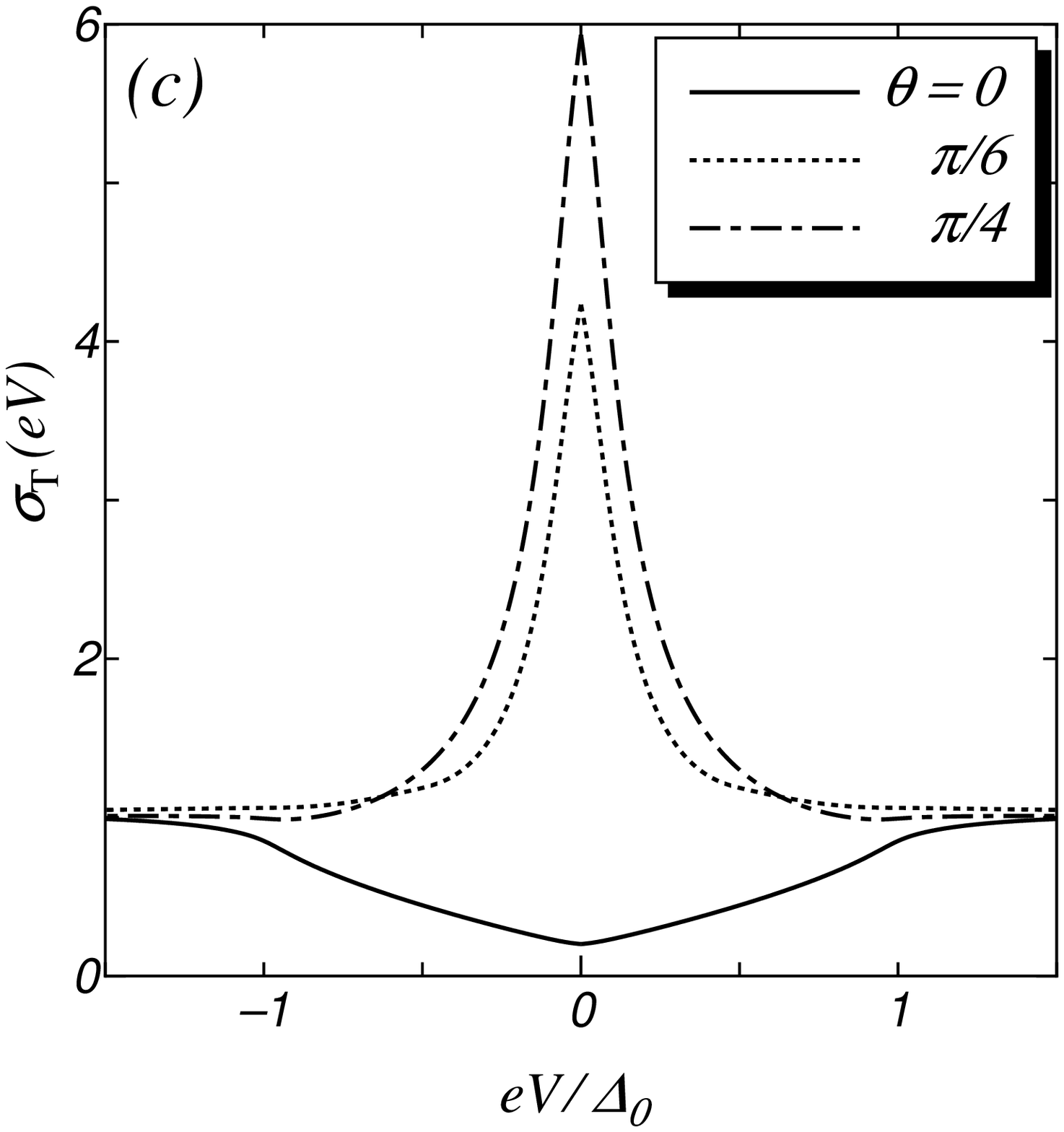}
\end{minipage}
\begin{minipage}[t]{8cm}
\epsfxsize=8cm
\epsfbox{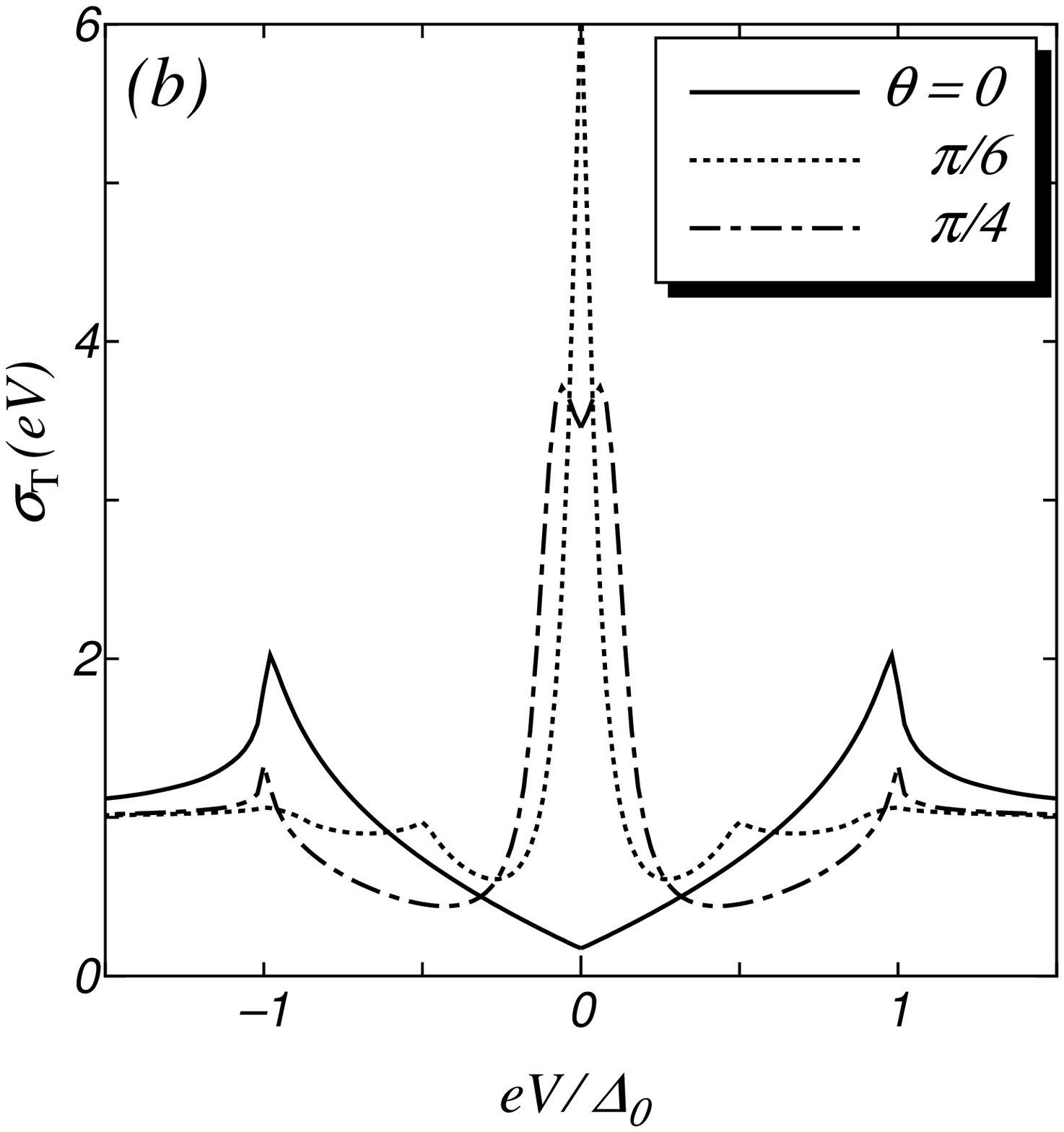}
\end{minipage}
\begin{minipage}[t]{8cm}
\epsfxsize=8cm
\epsfbox{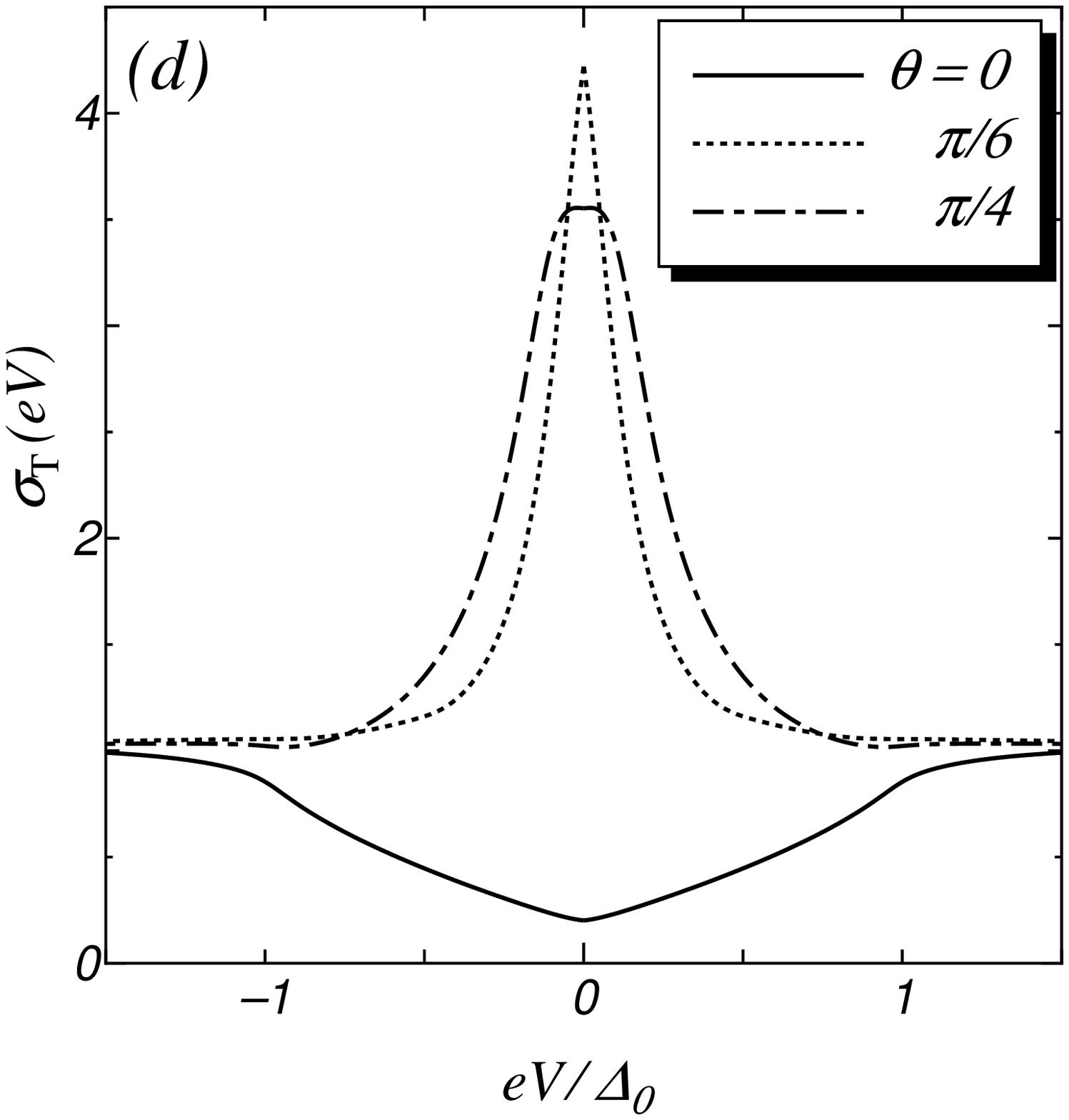}
\end{minipage}
\end{center}
\caption{
The normalized tunneling conductance
for $d_{x^{2}-y^{2}}$+i$d_{xy}$-wave state
with $Z=3.0$. 
(a) $T_{d_{xy}}/T_{d}=0.1$ and $T=0.0$. 
(b) $T_{d_{xy}}/T_{d}=0.2$ and $T=0.0$. 
(c) $T_{d_{xy}}/T_{d}=0.1$ and $T=0.05T_{c}$. 
(d) $T_{d_{xy}}/T_{d}=0.2$ and $T=0.05T_{c}$. 
\label{fig10}}
\end{figure}

%
The resulting normalized tunneling conductance
$\sigma_{\rm T}(eV)$
is plotted in 
Figs.~\ref{fig10}(a) and \ref{fig10}(b)
with $T=0$ and 
the corresponding quantities with $T=0.05T_{c}$ 
is plotted in 
Figs.~\ref{fig10}(c) and \ref{fig10}(d) for various $\theta$.
Only for $\theta=0$,
line shape of $\sigma_{\rm T}(eV)$
is similar to that of the bulk density of states
of $d_{x^{2}-y^{2}}$-wave superconducting state. 
In other cases, 
$\sigma_{\rm T}(eV)$ has a zero-bias enhanced line shape 
due to the formation of ABS [see Fig.~\ref{fig04}(a)].  
The ABS causes the ZBCP
when the magnitude of $T_{d_{xy}}$ is small.
However, with the increase of the magnitude of $T_{d_{xy}}$,
the zero-energy ABS is unstable and subdominant
$d_{xy}$-wave component is induced.
The energy levels of  bound state  shifts from zero
and the local density of states has a zero-energy peak splitting.
The resulting $\sigma_{\rm T}(eV)$ has a
ZBCP splitting as shown in Fig.~\ref{fig10}(b).
As compared to the case of $d_{x^{2}-y^{2}}$+i$s$ state,
the magnitude of $\sigma_{\rm T}(0)$ is not reduced seriously
since the effective $d_{xy}$-wave pair potential
felt by quasiparticles is distributed from
$-\mbox{Im}[\Delta_{d_{xy}}(0)]$ to
$\mbox{Im}[\Delta_{d_{xy}}(0)]$.
With the increase of  temperature,
the ZBCP splitting is not visible any more.
[see  Fig.~\ref{fig10}(c) and Fig.~\ref{fig10}(d)]. 
\par
\begin{figure}[htbp]
\begin{center}
\epsfxsize=15cm
\epsfysize=18.5cm
\centerline{\epsfbox{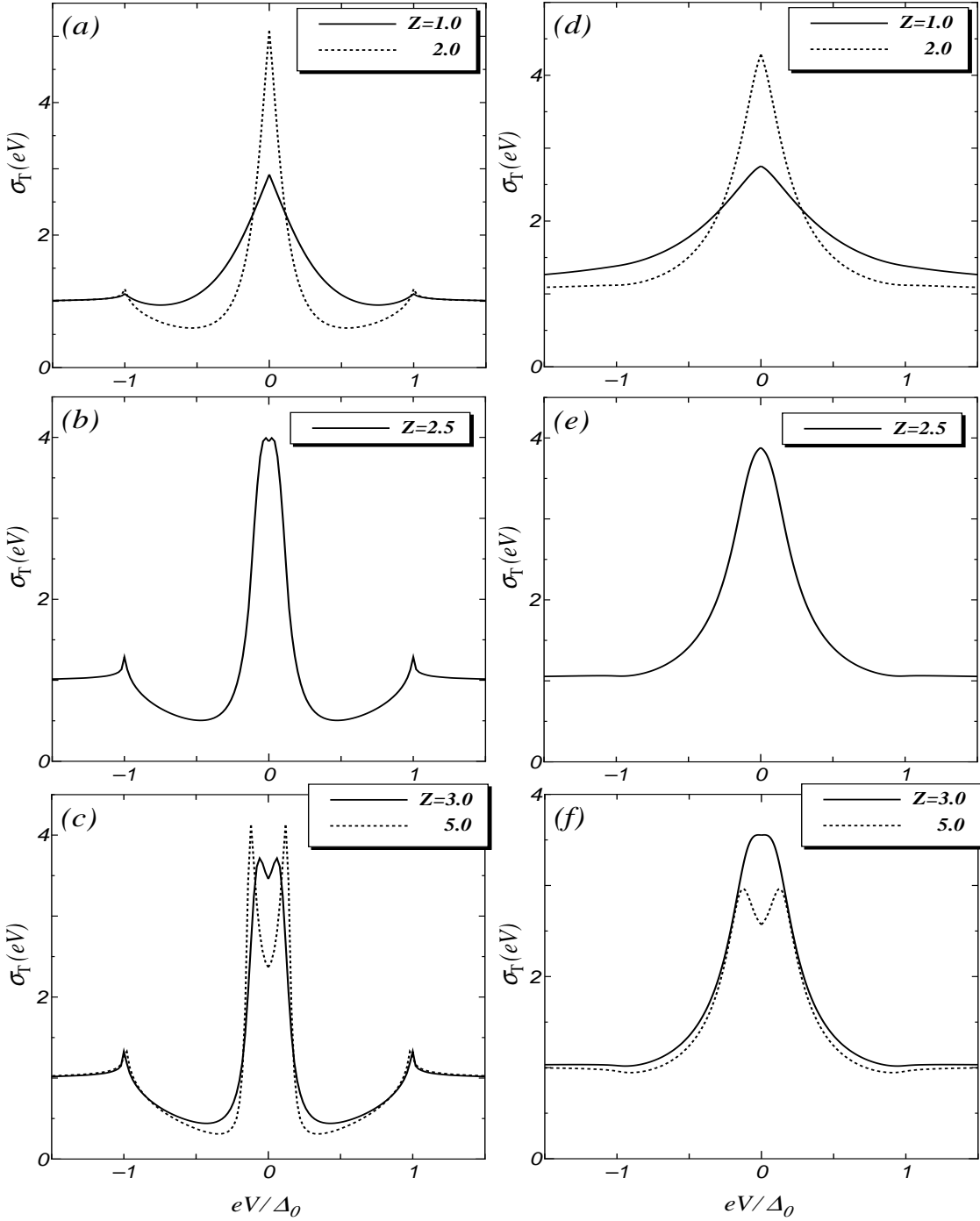}}
\caption{
Tunneling conductance 
for $d_{x^{2}-y^{2}}$+i$d_{xy}$-wave state
with $\theta=\pi/4$ and $T_{d_{xy}}/T_{d}=0.2$
for various $Z$. 
(a) [(d)] low barrier ($Z=1.0$, $2.0$)
(b) [(e)] middle barrier ($Z=2.5$),
and
(c) [(f)] high barier ($Z=3.0$, $5.0$).
$T=0$ for (a), (b), and (c). 
$T=0.05T_{c}$ for (d), (e), and (f). 
\label{fig11}}
\end{center}
\end{figure}

Next, we  concentrate on 
how $\sigma_{\rm T}(eV)$ is influenced
by the transmissivity of the junctions,
$i.e.$, the magnitude of $Z$.
In Fig.~\ref{fig11}, $\sigma_{\rm T}(eV)$ with $\theta=\pi/4$
is plotted for $T=0$ (left panels)
and $T=0.05T_{c}$ (right panels).
As shown in Fig.~\ref{fig11},
$\sigma_{\rm T}(eV)$ with $\theta = \pi/4$
depends on the transmissivity of the junction crucially.
For the junctions with high transmissivity, 
$\sigma_{\rm T}(eV)$ has a ZBCP [see Fig.~\ref{fig11}(a)]
and the magnitude of $\sigma_{\rm T}(0)$
is firstly enhanced with the increase of $Z$.
In this case, only the predominant
$d_{x^{2}-y^{2}}$-wave component
exists near the interface
as shown in Fig.~\ref{fig01}(b).
However, with the increase of $Z$,
$\sigma_{\rm T}(eV)$ starts to have
a peak splitting at a certain value of $Z$,
where the magnitude of the induced subdominant 
component at the interface 
$\mbox{Im}[\Delta_{d_{xy}}(0)]$
overlaps that of the predominant one
$\mbox{Re}[\Delta_{d}(x)]$.
For sufficiently larger magnitude of $Z$,
$\sigma_{\rm T}(eV)$ has a  ZBCP splitting 
[see Figs.~\ref{fig11}(b) and \ref{fig11}(c)]
and the magnitude of $\sigma_{\rm T}(0)$ 
decreases with the increase of $Z$.
The right panels of Fig.~\ref{fig11}
are shown for the $\sigma_{\rm T}(eV)$
with finite temperature.
Only for the junctions with the lowest transmissivity ($Z=5$),
the subtle splitting of ZBCP appears in $\sigma_{\rm T}(eV)$.
For $T=0.05T_{C}$, similar line shapes of $\sigma_{T}(eV)$ 
with smeared ZBCP splitting are obtained 
[see Fig.~\ref{fig11}(d), Fig.~\ref{fig11}(e) and 
Fig.~\ref{fig11}(f)]. 
However, the slight enhanced structure of 
$\sigma_{T}(eV)$ at $eV=\pm \Delta_{0}$ vanishes. 
\par
\begin{figure}[htbp]
\begin{center}
\epsfxsize=7cm
\epsfysize=18cm
\centerline{\epsfbox{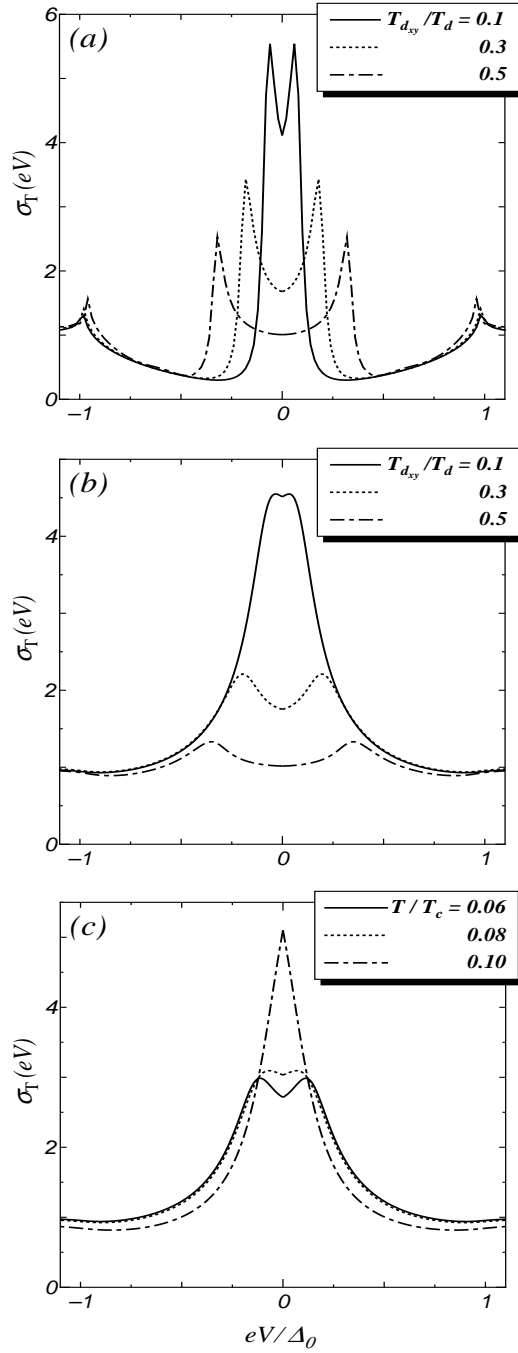}}
\caption{
Tunneling conductance
for $d_{x^{2}-y^{2}}$+i$d_{xy}$-wave state
with $\theta=\pi/4$ and $Z=5$. 
(a) $T=0$, 
(b) $T=0.05T_{c}$ 
for various  magnitude of $T_{s}$. 
(c)  various temperature with $T_{s}/T_{d}=0.2$.
\label{fig12}}
\end{center}
\end{figure}

Finally, we  look at the relation between the 
position of the splitted peak and the 
magnitude of $\mbox{Im}[\Delta_{d_{xy}}(0)]$.
In Fig.~\ref{fig12}(a), we show $T_{d_{xy}}$ 
dependence of the tunneling conductance spectra 
$\sigma_{\rm T}(eV)$ for $\theta=0$ with $Z=5$.
As shown in Fig.~\ref{fig09}(b), 
the magnitude of the induced subdominant imaginary 
component of $\mbox{Im}[\Delta_{d_{xy}}(0)]$ is
about $0.15\Delta_{0}$, $0.35\Delta_{0}$, and 
$0.52\Delta_{0}$ 
for $T_{d_{xy}}/T_{d}=0.1$, $T_{d_{xy}}/T_{d}=0.3$ 
and $T_{d_{xy}}/T_{d}=0.5$, respectively. 
The corresponding $\sigma_{\rm T}(eV)$ has a splitted peak
locating  at $\pm 0.16\Delta_{0}$, $\pm 0.3\Delta_{0}$,
and $\pm 0.42\Delta_{0}$, respectively. 
The width of the two splitted peaks 
depends on the temperature 
as shown in Fig.~\ref{fig12}(b).
Next, we show how the line shape of
$\sigma_{\rm T}(eV)$ changes at the temperature $T=T_{\bar{d_{xy}}}$
[$T_{\bar{d_{xy}}}=0.07T_{c}$] 
where  $\mbox{Im}[\Delta_{d_{xy}}(x)]$  becomes nonzero. 
As seen from Fig.~\ref{fig12}(c), the magnitude of 
$\sigma_{\rm T}(0)$ is reduced with the introduction of 
$\mbox{Im}[\Delta_{s}(x)]$. 
\par
At the end of this section, 
we can summarize that 
even in the presence of BTRSS, 
the resulting $\sigma_{\rm T}(eV)$
does not always have a clear ZBCP splitting 
due to the finite temperature effect 
when the transmission probability of the particles
at the interface is not low.
In order to observe the  ZBCP splitting which is 
one of the evidence supporting BTRSS, 
we must measure
$\sigma_{\rm T}(eV)$ for the junctions
with low transmissivity
with (110) oriented interface ($\theta=\pi/4$) 
at low temperatures. 
In such a case, 
we can classify the 
the $d_{x^{2}-y^{2}}$+i$s$- and
$d_{x^{2}-y^{2}}$+i$d_{xy}$-wave state
through the magnitude of $\sigma_{\rm T}(0)$ 
and the width of the  ZBCP splitting 
for the junctions with low transmission probability. 
\par
%
\section{Conclusion}
\label{sec:4}
In this paper, spatial dependence
of the pair potential in the N/D junctions is
determined on the basis of the quasiclassical theory
in the presence of subdominant component of the pair potential 
near the interface.
We clarified the influence of
the spatial variation of the pair potentials
on the tunneling conductance spectra
for various conditions of the junctions.
We selected two kinds of subdominant components
$s$- and $d_{xy}$-wave which are 
induced as $d_{x^{2}-y^{2}}$+i$s$- and
$d_{x^{2}-y^{2}}$+i$d_{xy}$-wave state, respectively.
The amplitude of
$\Delta_{s}(x)$ [$\Delta_{d_{xy}}(x)$] is enhanced
with the increase of the 
magnitude of $Z$, $T_{s}[T_{d_{xy}}]$, and $\theta$
($0<\theta <\pi/4$).
In the intermediate $\theta$,
both $\mbox{Im}[\Delta_{d}(x)]$ 
and $\mbox{Re}[\Delta_{s}(x)]$
$\{ \mbox{Re}[\Delta_{d_{xy}}] \}$
becomes nonzero,
although these magnitudes are small
as compared to those of $\mbox{Re}[\Delta_{d}(x)]$
and $\mbox{Im}[\Delta_{s}(x)]$ $\{ \mbox{Im}[\Delta_{d_{xy}}] \}$.
The resulting $\sigma_{\rm T}(eV)$ depends on 
$Z$, $T_{s}$ (or $T_{d_{xy}}$), and $\theta$. 
For fixed $Z$, the magnitude of $\sigma_{\rm T}(0)$
increases with the increase of $\theta$ $(0<\theta<\pi/4)$
for small magnitude of $T_{s}/T_{d}$ [$T_{d_{xy}}/T_{d}$].
While $\sigma_{\rm T}(0)$ increases and decreases again
with the increase of $\theta$ $(0<\theta<\pi/4)$
due to the formation of 
$d_{x^{2}-y^{2}}$+i$s$-
[$d_{x^{2}-y^{2}}$+i$d_{xy}$]-wave state for sufficiently larger
magnitude of $T_{s}/T_{d}$ [$T_{d_{xy}}/T_{d}$].
For fixed $\theta$ and $T_{s}/T_{d}$ [$T_{d_{xy}}/T_{d}$],
$\sigma_{\rm T}(eV)$ has a ZBCP for small magnitude of $Z$,
while it has a ZBCP splitting for sufficiently 
larger magnitude of $Z$.
Using junctions with small transmissivity,
we can distinguish
$d_{x^{2}-y^{2}}$+i$s$-wave state from
$d_{x^{2}-y^{2}}$+i$d_{xy}$-wave state 
since the magnitude of $\sigma_{\rm T}(0)$
for $d_{x^{2}-y^{2}}$+i$s$-wave state is much more reduced
as compared to that for $d_{x^{2}-y^{2}}$+i$d_{xy}$-wave state
as seen from Fig.~\ref{fig04} [Fig.~\ref{fig06}]
to Fig.~\ref{fig10} [Fig.~\ref{fig12}]. 
By taking into account finite temperature effect,
the degree of the ZBCP splitting is suppressed 
and the fine structure at $eV \sim \Delta_{0}$
in $\sigma_{\rm T}(eV)$ becomes invisible.
\par
In the light of our theory,
one of the reason for the absence of ZBCP
splitting for many tunneling experiments
may be due to their high transmissivity of the junctions 
(small magnitude of $Z$) and high temperatures.
If we choose $T_{c}$ as 90K, 
unless the transmissivity of the junction 
is sufficiently low, it is difficult to observe 
ZBCP splitting clearly at 4.5K. 
In order to see the ZBCP splitting clearly, 
we must observe much more low temperature 
($T<4.5K$) 
using a clean junctions  with low transmissivity 
for $\theta=\pi/4$. 
\par
In this paper,
we have chosen a free electron model
with cylindrical Fermi surfaces.
In order to compare actual tunneling conductance,
we must calculate spatial dependence of the pair potential
and the tunneling conductance
by taking into account of the actual shape of Fermi surface.
For this purpose, it is a promising way
to perform the calculation based on the 
$t$-$J$ model with Gutzwiller approximation \cite{Yokoyama},
where the doping dependence is naturally taken into account.
Using this model,
we succeeded to explain detailed line shape of
the scanning tunneling conductance spectra
around Zn impurity \cite{Pan,Tsuchiura1,Tsuchiura2}.
As regards quasiparticle states near the  interface, 
the pre-existing calculations are performed 
for infinite barrier limit \cite{Tanuma1,Tanuma2,Zhu}. 
In order to compare actual tunneling experiments,
we must calculate the N/D junctions based on the
$t$-$J$ model using Gutzwiller approximation.
In the $t$-$J$ model,
the subdominant $d_{xy}$-wave component
is hard to be realized \cite{Yokoyama}
since a nearest neighbor attractive potential is taken into account.
In order to discuss the $d_{xy}$-wave component,
we must consider much more long range interaction
including three site hopping term.
\par
It is also revealed by recent theory
that the BTRSS influences significantly
on the Josephson effect \cite{Tanaka96,Barash96,Tanaka98}.
The previous theory assumes a sufficient larger magnitude of 
barrier \cite{Tanaka98}.
In order to compare the existing experiments,
we must calculate Josephson current in the presence 
of BTRSS for arbitrary
transmissivity of the junctions.
\par
%
\section*{Acknowledgments}
First author (Y.T.) would like to thank
K. Machida and M. Ichioka
for useful and fruitful discussions.
He acknowledges the financial support of Research
Fellowships of Japan Society for the Promotion
of Science (JSPS) for Young Scientists.
This work is supported by
the Core Research for
Evolutional Science and Technology (CREST) of the Japan Science
and Technology Corporation (JST).
The computational aspect of this work has been performed at the
facilities of the Supercomputer Center, Institute for Solid State Physics,
University of Tokyo and the Computer Center.
%
%

\end{document}